\documentclass[twocolumn]{aastex62}

\graphicspath{{./}{figures/}}

\received{August 08, 2019}
\revised{September 12, 2019}
\accepted{September 17, 2019}

\submitjournal{ApJ}

\shorttitle{NTF Magnetism}
\shortauthors{Par\'e et al.}

\begin{document}

\title{A VLA Polarimetric Study of the Galactic Center Radio Arc: Characterizing Polarization, Rotation Measure, and Magnetic Field Properties}

\correspondingauthor{Dylan Par\'e}
\email{dylan-pare@uiowa.edu}

\author[0000-0002-5811-0136]{Dylan M. Par\'e}
\affil{University of Iowa \\
30 North Dubuque Street, Room 203 \\
Iowa City, IA 52242}

\author{Cornelia C. Lang}
\affil{University of Iowa \\
30 North Dubuque Street, Room 203 \\
Iowa City, IA 52242}

\author[0000-0002-6753-2066]{Mark R. Morris}
\affil{University of California, Los Angeles \\
430 Portola Plaza, Box 951547 \\
Los Angeles, CA 90095-1547}

\author{Hailey Moore}
\affil{University of Iowa \\
30 North Dubuque Street, Room 203 \\
Iowa City, IA 52242}

\author{Sui Ann Mao}
\affil{Max Planck Institute for Radio Astronomy \\
Auf dem $H\ddot{u}gel$, P.O. Box 20 24 \\
Bonn, Germany D-53010}

\begin{abstract}

The Radio Arc is one of the brightest systems of non-thermal filaments (NTFs) in the Galactic Center, located near several prominent HII regions (Sickle and Pistol) and the Quintuplet stellar cluster. We present observations of the Arc NTFs using the S-, C-, and X-bands of the Very Large Array interferometer. Our images of total intensity reveal large-scale helical features that surround the Arc NTFs, very narrow sub-filamentation, and compact sources along the NTFs. The distribution of polarized intensity is confined to a relatively small area along the NTFs. There are elongated polarized structures that appear to lack total intensity counterparts. We detect a range of rotation measure values from -1000 to -5800 rad m$\rm^{-2}$, likely caused by external Faraday rotation along the line of sight. After correcting for Faraday rotation, the intrinsic magnetic field orientation is found to generally trace the extent of the NTFs. However, the intrinsic magnetic field in several regions of the Arc NTFs shows an ordered pattern that is rotated with respect to the extent of the NTFs. We suggest this changing pattern may be caused by an additional magnetized source along the line of sight, so that we observe two field systems superposed in our observations. We suggest that the large scale helical segments near the Radio Arc could be components of such a source causing these changes in intrinsic magnetic field, and some variations in the polarization and rotation measure values along the NTFs. 

\end{abstract}


\section{INTRODUCTION} \label{sec:intro}

The Galactic Center (GC) region of the Milky Way is a galactic nuclear region only 8 kpc away (adopting the GC-Earth distance used in \citet{Abuter2019}). The proximity of the GC allows us to infer properties of more distant galactic nuclear regions that are Mpc away. The GC is a unique environment exhibiting properties that differ from the rest of the Galactic disk. For example, the central 200 pc of the Galaxy features multiple HII regions; high molecular densities ($\rm 10^3$ - $\rm 10^7\,cm^{-3}$); high molecular gas temperatures (50-300 K); several massive, young star clusters; and the presence of the supermassive black hole Sgr $\rm A^*$ \citep{Huettemeister1993,Figer1999,Mills2014,Mills2018,Abuter2018}. There is also a population of radio continuum structures appearing as filaments which seem unique to the GC \citep{Morris1996,Gray1995,LaRosa2006}.

These unusual radio filaments are highly polarized synchrotron emission sources appearing only throughout the central few degrees of the Galaxy \citep{YMC1984,Yusef-Zadeh1987a}. These non-thermal filaments (NTFs) are 10-40 pc in length and $\rm\lesssim$ 0.5 pc in width \citep{Morris2007}. Their synchrotron emission indicates that free electrons present in the GC are being accelerated to relativistic speeds in the presence of strong and ordered local magnetic fields \citep{YMC1984}, though general agreement has not been reached on either the electron source or mechanism of acceleration.

The NTFs are either flux tubes tracing a strong, pervasive magnetic field within the GC or local enhancements to a weaker GC magnetic field of 10s $\rm\mu$G  \citep{Uchida1996,Morris1996,Staguhn1998,Shore1999,Bicknell2001,Ferriere2009}. Determining which of these scenarios is more applicable will allow us to probe the structure of the magnetic field in the GC. Characterizing this structure will allow us to expand our understanding of galactic magnetism by supplementing previous efforts to analyze magnetic fields in galactic disks and halos \citep{Heald2009,Irwin2012a,Beck2013}.

The first detected system of NTFs in the GC is the Radio Arc (hereafter referred to as the ``Arc NTFs''), and it is the brightest NTF system in the GC. As such, the Arc NTFs can be used to infer properties of the NTF population as a whole. While the Arc NTFs possess a continuous total intensity distribution throughout their length, the corresponding polarized intensity is ``patchy'' or discontinuous, revealing discrete clumps of significant polarized intensity separated by regions of complete depolarization \citep{Yusef-Zadeh1986a,YM1987,Inoue1989}. Other filaments have been observed with similar polarized intensity distributions, indicating that this is a general feature of NTFs \citep{Yusef-Zadeh1989,YWP1997,Lang1999a,Lang1999b}. The Radio Arc is known to possess a flat spectral index \citep{Yusef-Zadeh1986a} and to consist of a number of closely packed NTFs \citep{YM1987}.

Rotation Measure (RM) distributions for the NTFs, which is the integral over the line of sight of the product of the line of sight magnetic field and electron density, have unveiled RM magnitudes as high as 5500 rad m$\rm^{-2}$ \citep{Yusef-Zadeh1986a,Yusef-Zadeh1987a}, but were limited by narrow spectral bandwidths. For example, \citet{Yusef-Zadeh1987a} were limited to only two wavelengths (6 and 20 cm) with bandwidths of at most 50 MHz when estimating the RM of the Arc NTFs \citep{YM1987}. Since the RM value is the slope of the best-fit model of the polarized intensity angle vs. wavelength squared, a larger frequency range studied allows for a more constrained determination of the RM value \citep{Burn1966}. An increased spectral resolution also reduces the impact of bandwidth depolarization. The updated resolution and spectropolarimetric capabilities of the NSF's Karl G. Jansky Very Large Array (VLA) allow us to build on the original observations of polarized intensity and RM distributions.

Because only a few RM studies have been conducted for the NTFs at the GC, there are not many examples of intrinsic magnetic field distributions for the NTFs. Three studies have corrected for the RM and obtained intrinsic magnetic field distributions for NTFs G358.85+0.47, G0.08+0.15, and G359.54+0.18 \citep{Lang1999a,Lang1999b,YWP1997}. In these NTFs, the intrinsic magnetic field orientations are primarily oriented along the long axis of the NTFs, consistent with the idea that these structures have a highly ordered field structure. For the Arc NTFs, the observed electric field vectors were presented in \citep{Yusef-Zadeh1986a} and are uncorrected for Faraday rotation. 

In this paper we study (1) the total intensity distribution of the Arc NTFs and surrounding environment, (2) the structure and frequency dependence of the polarized intensity of the Arc NTFs, (3) the magnitude and distribution of the RM toward the Arc NTFs, and (4) the intrinsic magnetic field of the Arc NTFs.

In Section \ref{sec:meth} we describe the observations and data reduction. Section \ref{sec:tot_res} presents the results of our total intensity observations whereas Section \ref{sec:pol_res} presents the polarization results. In Section \ref{sec:disc} we discuss and interpret our results and our conclusions are presented in Section \ref{sec:conc}.

\section{OBSERVATIONS AND DATA REDUCTION} \label{sec:meth}
	
\subsection{Observations} \label{sec:obs}

\begin{deluxetable*}{|c|c|c|c|c|c|}[ht!]
\tablecaption{Summary of VLA GC Radio Arc Observations}
\tablecolumns{6}
\tablenum{1}
\tablewidth{0pt}
\tablehead{
\colhead{Field} & \multicolumn{2}{c}{Pointing Center (J2000)}\tablenotemark{a} & \colhead{} & \colhead{Array} & \colhead{} \\
\colhead{Name} & \colhead{\textbf{RA}} & \colhead{\textbf{DEC}} & \colhead{Frequency (GHz)} & \colhead{Configuration} & \colhead{Integration Time (min)\tablenotemark{b}}
}
\startdata
S & 17:46:32.00 & -28:51:16.0 & 1.988 - 4.012 & DnC, CnB, BnA & 10, 47, 13 \\
C1 & 17:46:35.95 & -28:52:05.3 & 3.988 - 6.012 & DnC, CnB, B, BnA & 10, 12, 30, 12 \\
C2 & 17:46:21.96 & -28:49:55.4 & 3.988 - 6.012 & ... & 10, 12, 30, 11 \\
X1 & 17:46:41.70 & -28:53:21.9 & 9.976 - 12.000 & DnC, CnB, B & 10, 10, 33 \\
X2 & 17:46:32.63 & -28:51:49.6 & 9.976 - 12.000 & ... & 10, 10, 32 \\
X3 & 17:46:24.21 & -28:50:13.6 & 9.976 - 12.000 & ... & 11, 10, 32 \\
X4 & 17:46:15.20 & -28:48:45.5 & 9.976 - 12.000 & ... & 10, 10, 32 \\
\enddata
\tablenotetext{a}{Positional errors are $\rm\pm$0.01s in RA and $\rm\pm$0.1\arcsec~in DEC.}
\tablenotetext{b}{Integration times rounded to the nearest whole minute. Times are arranged to match the arrays used as in the array configuration column.}
\label{table:obs}
\end{deluxetable*}

The Radio Arc NTFs were observed with the VLA over 9 days in 2013$-$14. Multiple VLA configurations were used during these observations (see Table \ref{table:obs}), which provide sensitivity to both diffuse and compact structure.  Three VLA observing bands were used (S-, C- and X-bands) with central frequencies of 3.0, 5.0, and 10.9 GHz respectively. Our data span a large frequency range (approximately 2-12 GHz, not contiguous) which is sampled every 8 MHz using the upgraded capabilities of the VLA, providing hundreds of channels and the ability to do spectro-polarimetric analysis for the first time using the images of the Arc NTFs. 

\subsection{Calibration} \label{sec:cal}

At each observing freqeuency, three calibrators were used: 3C286 (flux and bandpass), J1407+2827 (polarization leakage), and J1751-2524 (phase calibration). Calibration corrections to the raw measurement sets were applied using the Common Astronomy Software Applications (CASA) software. The standard calibration steps were performed with limited self-calibration used in situations where residual phase or amplitude errors were present. For the C-band data set a frequency range from 4.0$-$4.2 GHz was flagged due to transmissions from geosynchronous satellites. Similarly, for the X-band data set the 11.7$-$12.0 GHz frequency range was plagued by strong RFI, and so was completely flagged. Outside of these specific frequency ranges, only routine flagging procedures were employed. 

\subsection{Total Intensity Imaging} \label{sec:I_imag}

To produce the total intensity images (Stokes I) shown in Figures \ref{fig:tot_I_1}-\ref{fig:tot_I_4}, we used the CASA \textit{clean} routine with large numbers ($\rm{}\geq$100,000) of iterations to account for all the complex structure in our field of view. MFS cleaning was used to generate 2D images of the Arc NTFs, with the individual pointings in X- and C-bands (refer to Table \ref{table:obs}) mosaiked for the cleaning. These images were made with high resolution (robust=1), which resulted in synthesized beams of 5.3\arcsec~x 3.5\arcsec~at 3 GHz (Figure \ref{fig:tot_I_1}), 2.3\arcsec~x 1.2\arcsec~at 5 GHz (Figure \ref{fig:tot_I_2}) and 1.4\arcsec~x 0.7\arcsec~at 10 GHz (Figure \ref{fig:tot_I_3}). 

\subsection{Polarized Intensity Imaging} \label{sec:P_imag}

The polarized intensity distributions (P) were made from the Q and U distributions using the equation: $\rm P = \sqrt{Q^2 + U^2}$, which was accomplished using the CASA task \textit{immath}. Stokes Q and U images were made for the data at 5 and 10 GHz using \textit{clean} with $\sim$100 iterations. Due to the large field of view at 3 GHz and challenges with correcting for off-axis polarization calibration, images of polarized intensity were not generated for the 3 GHz data. 

We generated two sets of polarized intensity distributions: (1) high-resolution images used for studying structure and substructure and (2) lower-resolution (smoothed) images for studying the distribution of the RM. The high-resolution images were made using the full resolution of the data and the resulting images of linear polarization distribution at 5 and 10 GHz are presented in Figure \ref{fig:ps}. The polarized intensity images were then thresholded to 3 times the rms noise level in the Q and U distributions, a value of 0.264 mJy beam$\rm^{-1}$ at 10 GHz and 0.177 mJy beam$\rm^{-1}$ at 5 GHz. 

%\subsubsection{Images for Rotation Measure Analysis} \label{sec:B_imag}

The lower-resolution polarized intensity images were used for the analysis of the RM values towards the Arc NTFs. For this analysis, we need to use as large a frequency range as possible (spanning the full 8 GHz range of the two bands), so the polarization images were smoothed to the lowest resolution in this range, i.e., 7\arcsec. In addition, we are interested in obtaining RM and intrinsic magnetic field results with a high signal-to-noise ratio, so we only use the DnC array data for the 5 and 10 GHz sets. Therefore, a second set of polarized intensity images was produced with a synthesized beam of 7\arcsec{} x 7\arcsec{}. The low-resolution polarized intensity distributions were then made from these smoothed linear polarization products using \textit{immath}. No debiasing was applied to these lower resolution polarized intensity images. 

The 10 GHz images for both data sets consisted of 208 channels and the 5 GHz images consisted of 224 channels. The channel widths were set to 8 MHz to account for small-scale variations between adjacent 2 MHz channels. Channels at the edge of the spectral windows (spws) in our data had low signal-to-noise ratios, and so these were removed before analysis of the data. This filtering process resulted in several 8 - 16 MHz gaps within both the 5 and 10 GHz data sets.

In order to solve for the RM toward a source, it is necessary to track the behavior of the polarization angle as a function of wavelength squared. The polarization angle ($\rm\chi$) at a given wavelength (or frequency) can be expressed as:
\begin{equation}
	\rm\chi = \frac{1}{2}\arctan\left(\frac{U}{Q}\right)
\end{equation}
where $\rm Q$ and $\rm U$ are the linear polarization products already introduced earlier in this section.

$\rm\chi$ distributions were made from the smoothed Q and U distributions using the \textit{immath} task. Our final product for the RM analysis is a cube that contains measurements of the polarization angle at 7\arcsec~resolution across $\rm\sim$ 200 frequency channels.

\section{TOTAL INTENSITY RESULTS} \label{sec:tot_res}

\begin{figure*}
    \centering
    \gridline{\fig{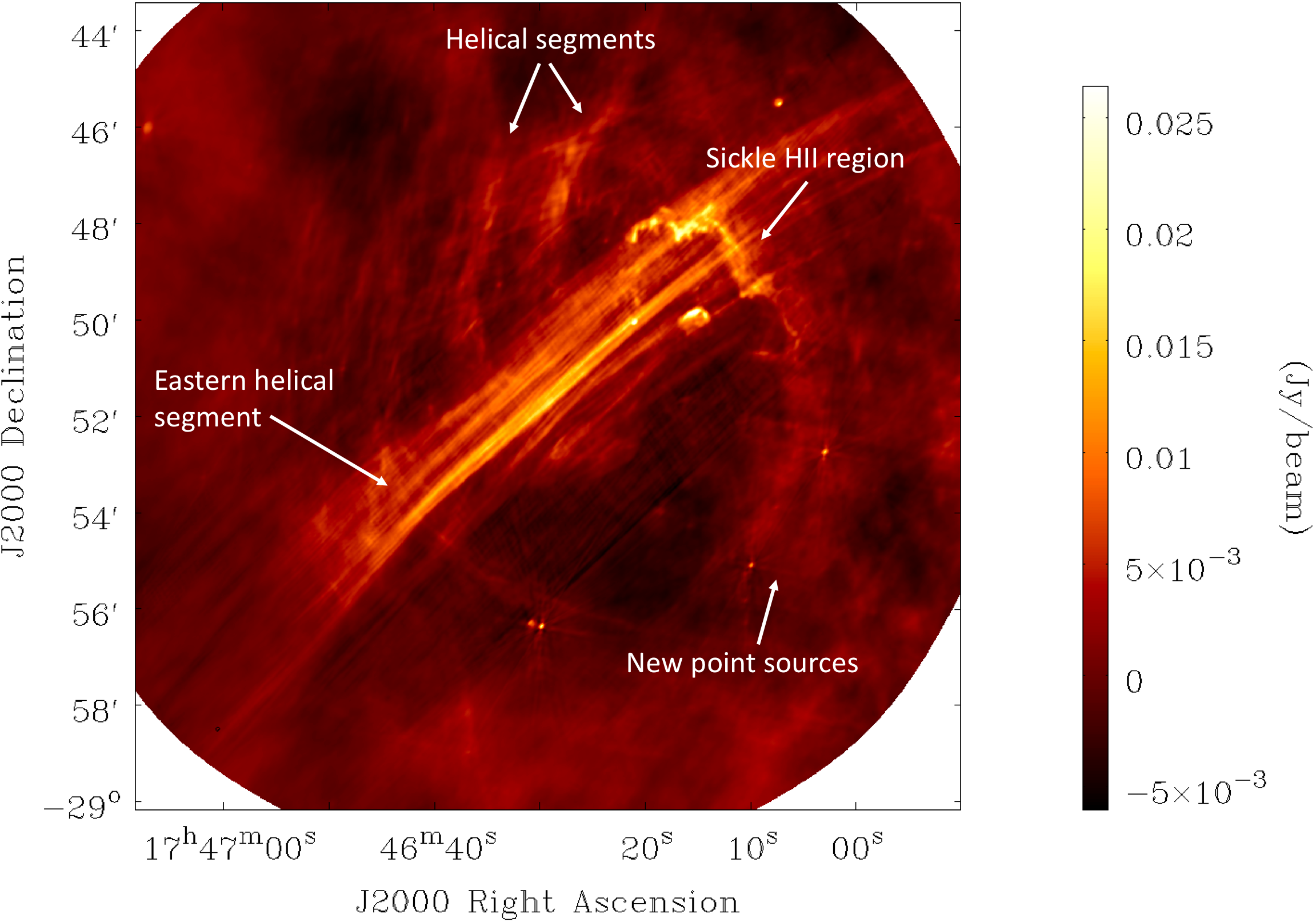}{0.9\textwidth}{}}
    \caption{3 GHz total intensity image of the Arc NTFs and surrounding region. Imaging was done over a 2 GHz frequency range from 2.0 - 4.0 GHz with no significant frequency gaps. The resolution of this image is 5.3\arcsec~x 3.5\arcsec~and the rms noise is 30 $\rm\mu$Jy beam$\rm^{-1}$. Features discussed in the text are marked with white arrows.}
    \label{fig:tot_I_1}
\end{figure*}

\begin{figure*}
    \centering
    \gridline{\fig{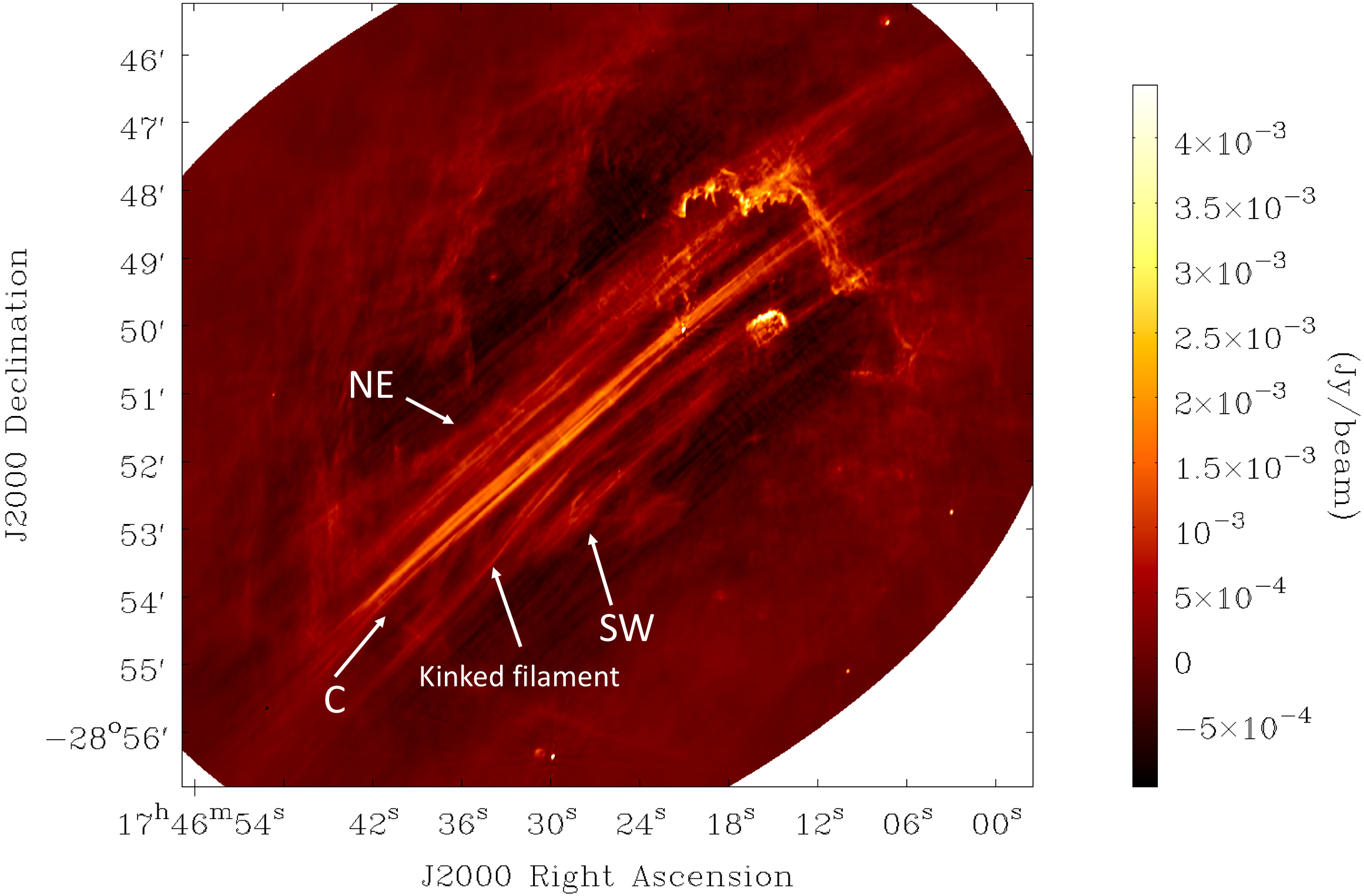}{0.9\textwidth}{}}
    \caption{5 GHz total intensity image of the Arc NTFs and surrounding region. Imaging was done over a frequency width of 4.3 - 5.9 GHz. The resolution of this image is 2.3\arcsec~x 1.2\arcsec~and the rms noise level is 8 $\rm\mu$Jy beam$\rm^{-1}$. Features discussed in the text are marked with white arrows.}
    \label{fig:tot_I_2}
\end{figure*}

\begin{figure*}
    \centering
    \gridline{\fig{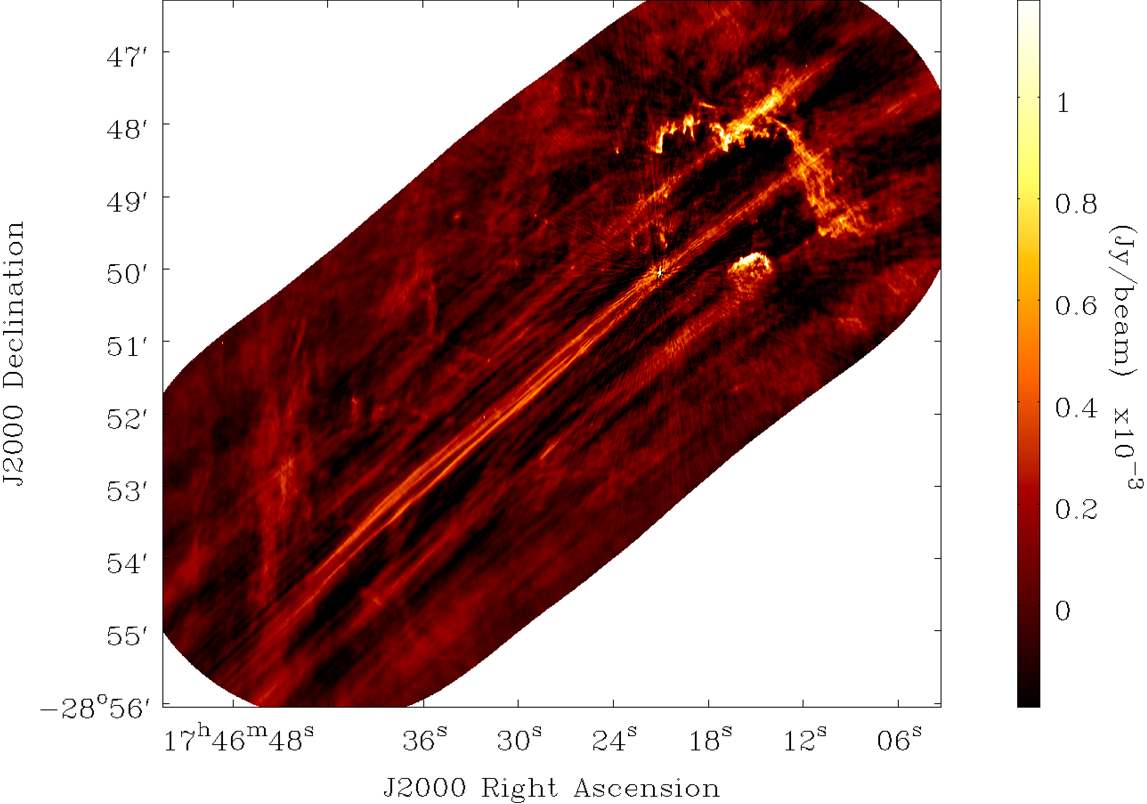}{0.9\textwidth}{}}
    \caption{10 GHz total intensity image of the Arc NTFs. Imaging was done over a frequency width of 10.0 - 11.7 GHz. The resolution of the image is 1.4\arcsec~x 0.7\arcsec~and the rms noise level is 15 $\rm\mu$Jy beam$\rm^{-1}$.}
    \label{fig:tot_I_3}
\end{figure*}

\begin{figure*}
    \centering
    \gridline{\fig{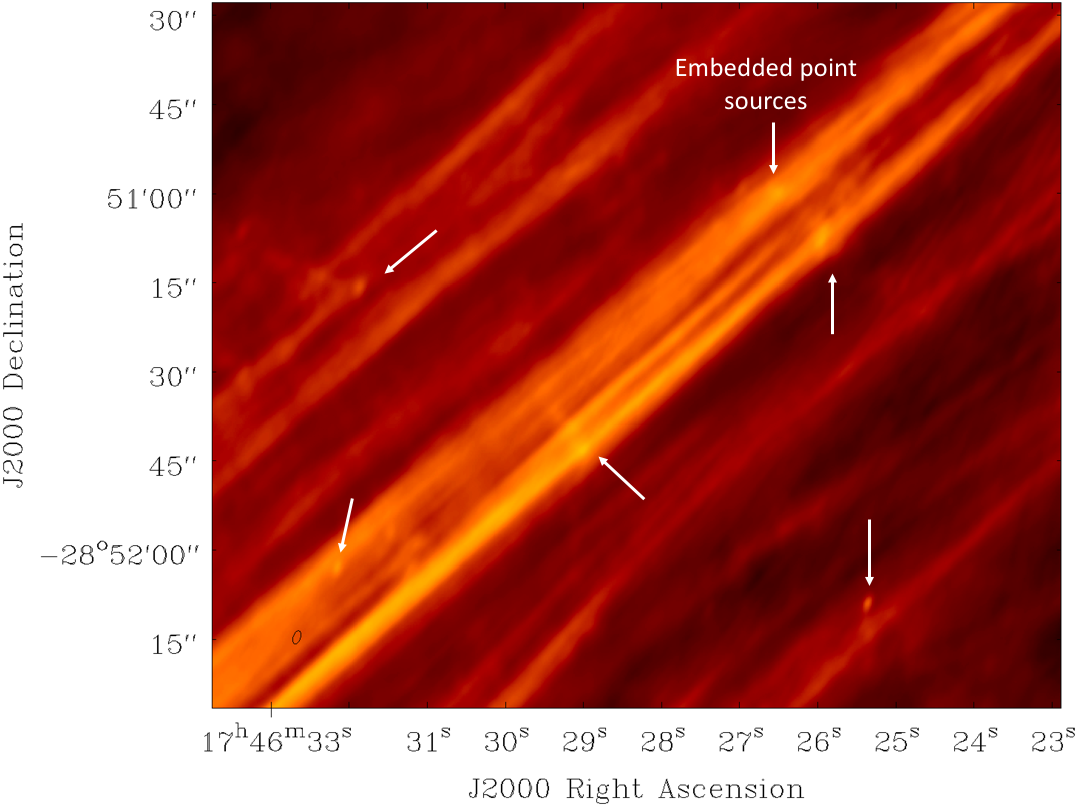}{0.4\textwidth}{}
              \fig{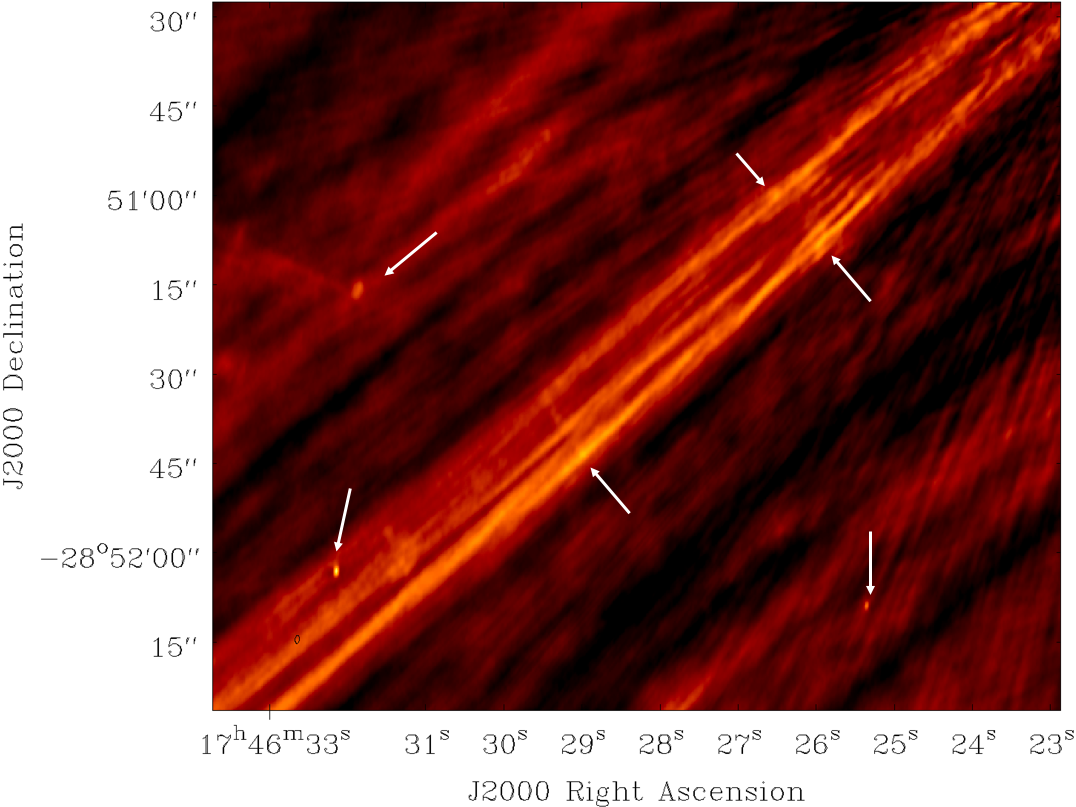}{0.4\textwidth}{}}
    \caption{A zoomed in region of the Arc NTFs at 5 GHz (left) and 10 GHz (right). Image parameters are the same as in Figures \ref{fig:tot_I_2} and \ref{fig:tot_I_3}. The sub-filamentation within the NTF bundles is clear in these panels. Also marked are a number of point sources seemingly embedded in the Arc NTFs.}
    \label{fig:tot_I_4}
\end{figure*}

\subsection{Overview of Total Intensity} \label{Sec:I_struct}
Figures \ref{fig:tot_I_1}-\ref{fig:tot_I_4} present total intensity images of the Arc NTFs at 3 GHz (Figure \ref{fig:tot_I_1}), 5 GHz (Figure \ref{fig:tot_I_2}) and 10 GHz (Figure \ref{fig:tot_I_3}). Figure 4 shows a zoomed-in region of Figures \ref{fig:tot_I_2} (left panel) and \ref{fig:tot_I_3} (right panel). The region shown in Figures \ref{fig:tot_I_1}-\ref{fig:tot_I_3} includes the entire Radio Arc complex: the thermal Sickle and Pistol HII regions \citep{Lang1997}, the Quintuplet Cluster \citep{Figer1999}, the non-thermal filaments (NTFs) \citep{Yusef-Zadeh1987a,YM1987} and the bright N3 point source located along the Arc NTFs \citep{Ludovici2016}. Owing to the improved sensitivity of the VLA, these images provide an even deeper and more sensitive look at the unique and unusual structures that pervade this region of the GC. The rms noise levels in these images are on the order of 10 $\rm\mu$Jy compared with 100s of $\rm\mu$Jy in previous studies. This improvement allows us to uncover new and intriguing features in the images of varying field of view and resolution. Figures \ref{fig:tot_I_1} - \ref{fig:tot_I_3} illustrate that the radio emission from the Radio Arc region is rich and complex in structure, including a mix of diffuse and narrow filamentary structures, bright and compact point-like sources and a variety of shell-like compact regions. 

As seen in Figure \ref{fig:tot_I_1}, the Arc NTFs are oriented nearly perpendicular to the orientation of the Galactic plane and are strikingly linear in their morphology over their full extent.  The prominent NTFs span the field of view for at least 15$\arcmin$ (35 pc at the GC distance of 8.0 kpc). Toward the southeastern edge of the field of view, the radio continuum brightness of the Arc NTFs becomes more diffuse and the individual NTFs are not as apparent. This decrease in NTF brightness is also observed in \citet{Yusef-Zadeh1987a} However, this drop in intensity and structure observed may be due to the primary beam coverage at 3 GHz and not inherent to the NTFs. In fact, VLA studies of the Arc NTFs at higher declinations \citep{Yusef-Zadeh1988}, as well as single dish studies \citep{Tsuboi1995} show the continuation of prominent NTFs beyond the primary beam of the 3 GHz image.  

Apparent to the north of the Arc NTFs in Figure \ref{fig:tot_I_1} are the ``helical segments'' first detected and discussed in  \citet{Yusef-Zadeh1987a,YM1987,Inoue1989}. These two curved, diffuse segments (centered on RA=17h 46m 30s and 35s, DEC=-28d 47\arcmin) may be part of a much larger structure that may wrap around the NTFs. This larger structure would have a helical shape, but we are only able to resolve segments of it due to only being able to resolve out to a certain largest angular size in our VLA observations. Thanks to our high sensitivity observations of the Radio Arc Complex, we obtain a more detailed view of these segments in Figures \ref{fig:tot_I_1} and \ref{fig:tot_I_2} than previous observations. The easternmost helical segment clearly passes through the Arc NTFs, and where it does a number of small, thread-like features appear to bridge adjacent NTFs (RA=17h 46m 50s, DEC=-28d 54\arcmin). Finally, the easternmost helical segment appears to be part of a large shell-like feature (best observed in Figures \ref{fig:tot_I_1} and \ref{fig:tot_I_2}).

\subsubsection{Filamentary Structure Within the Radio Arc NTFs}

Toward the center of Figure \ref{fig:tot_I_1} the NTFs have more apparent substructure. It is clear that the NTFs are comprised of multiple filaments: Figures \ref{fig:tot_I_3} and \ref{fig:tot_I_4} illustrate the complexity of the filamentary structure in the Radio Arc. The most striking features of the Arc NTFs are their multiplicity and their narrow widths. Unlike the rest of the prominent NTFs in the GC region, the Arc NTFs are comprised of many “bundles” of NTFs. In particular, we identify three main bundles of NTFs in the Radio Arc ordered as a function of longitude: the northern Arc NTFs, the Central Arc NTFs, and the southern Arc NTFs (labeled `NE,' `C,' and `SW' in Figure \ref{fig:tot_I_2}). In Figure \ref{fig:tot_I_1}, it is difficult to distinguish between these portions of the Radio Arc because of the brightness of all the NTFs. In Figures \ref{fig:tot_I_2} and \ref{fig:tot_I_3}, the central Arc NTF bundle appears significantly brighter along its length than the northern or southern Arc NTFs. 

The high resolution of these images allows many NTFs to be revealed. In fact, throughout the NTF bundle, more than 20 individual and primarily parallel NTFs can be identified. The high resolution of the images shows that the NTFs are very narrow, with the narrowest widths of $\sim$5\arcsec~(or 0.2 pc). Figures \ref{fig:tot_I_2} and \ref{fig:tot_I_3} reveal that the NTFs do not appear to twist along their lengths as a previous study had considered \citep{YM1987}; rather, the lower resolution previous images were not able to detect the fine scale structure that our data show. In addition to the very narrow widths of the NTFs, there are a number of bright and very compact point sources apparent within the NTF structure (of order the same size as the narrowest NTF width, i.e., $\sim$5\arcsec), as if they were “embedded” in the complex (especially apparent in Figures \ref{fig:tot_I_3} and \ref{fig:tot_I_4}). 

\subsubsection{New Filamentary Structures Near the Radio Arc NTFs} 

The improved sensitivity of the VLA observations we present here reveals a number of new filamentary features in the region surrounding the Radio Arc. In particular, we point out two interesting filamentary features. (1) Figure \ref{fig:tot_I_2} shows a linear filament (labeled in that figure) that appears to not remain parallel to the rest of the Arc NTFs; instead the filament has a different angle or may exhibit a kink near RA=17h 46m 36s DEC=-28d 54\arcmin. Other NTFs have been found to exhibit kink-like morphology along their lengths, including the ‘Snake’ \citep{Gray1995} and the ‘Pelican’ \citep{Lang1999a} and one idea is that a kink may occur during an interaction with an adjacent molecular cloud \citep{Yusef-Zadeh1987a,Tsuboi1997}. (2) Figure \ref{fig:tot_I_1} shows the helical segments originally detected by \citet{Morris2005} that seem to connect to the Radio Bubble seen in \citet{Simpson2007}. They are oriented neither parallel nor perpendicular to the Arc NTFs.

\subsubsection{Bright Compact Sources}

%The images presented in Figures \ref{fig:tot_I_1}-\ref{fig:tot_I_3} represent some of the highest sensitivity and highest resolution observations to date and allow us to detect bright compact sources that were undetected in previous studies of this region. 
The observed field contains numerous compact sources, of which only a few have been noted in previous studies: N3 (G0.17-0.08), N1 (G0.10-0.05), and O1 (G0.21- 0.00) (e.g., \citet{Yusef-Zadeh1986a,Yusef-Zadeh1987,Lang1997,Ludovici2016}). 

A set of bright compact sources are evident in Figure \ref{fig:tot_I_1}. A double lobed source that is located at the southern-most extent of the curved filament is a known maser \citep{Guesten1983}. Numerous other sources are visible, some previously unidentified ones which are labelled in Figure \ref{fig:tot_I_1}. Specfically the two bright point sources shown to the south-west of the Arc NTFs in Figure \ref{fig:tot_I_1} have not been identified previously.

In total, we have identified approximately 23 bright compact sources in Figures \ref{fig:tot_I_1}-\ref{fig:tot_I_4}. Many of these sources are resolved and show substructure, such as shell-like morphology or head-tail morphology associated with ultra-compact HII regions. Some of the sources are also unresolved. Rough estimates of spectral indices show a relatively equal distribution of thermal and non-thermal sources, though the background levels in this region make accurate spectral index calculations challenging.  A more detailed study of these point sources is deferred to a subsequent paper.

\section{POLARIZATION AND ROTATION MEASURE RESULTS} \label{sec:pol_res}
\begin{figure*}[htp]

\gridline{\fig{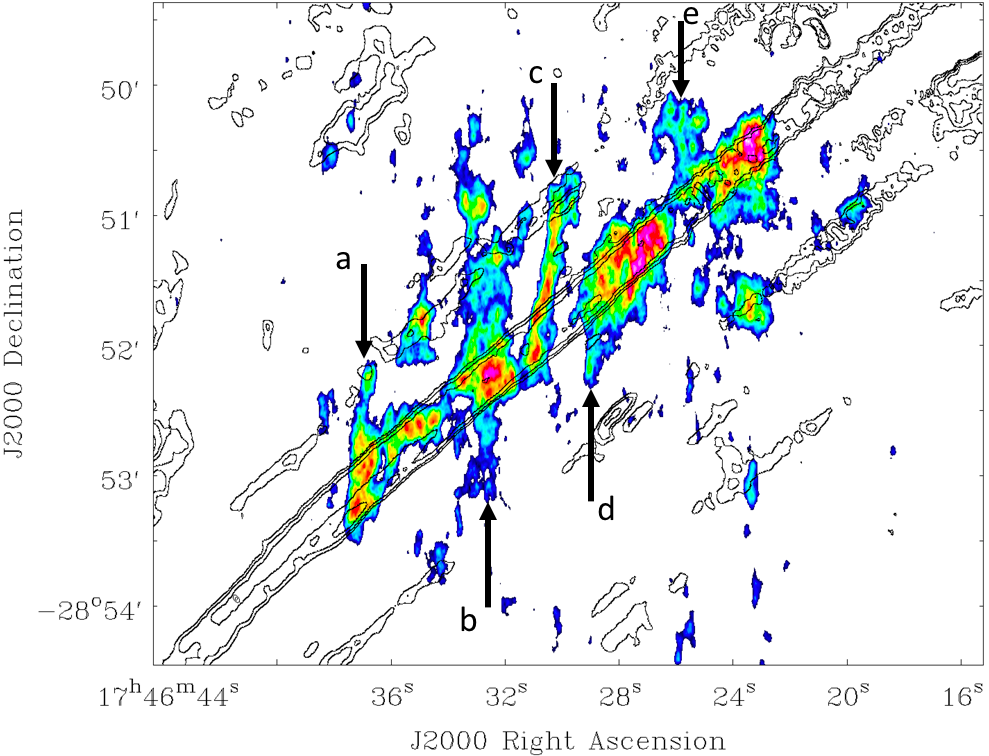}{0.7\textwidth}{}}
\gridline{\fig{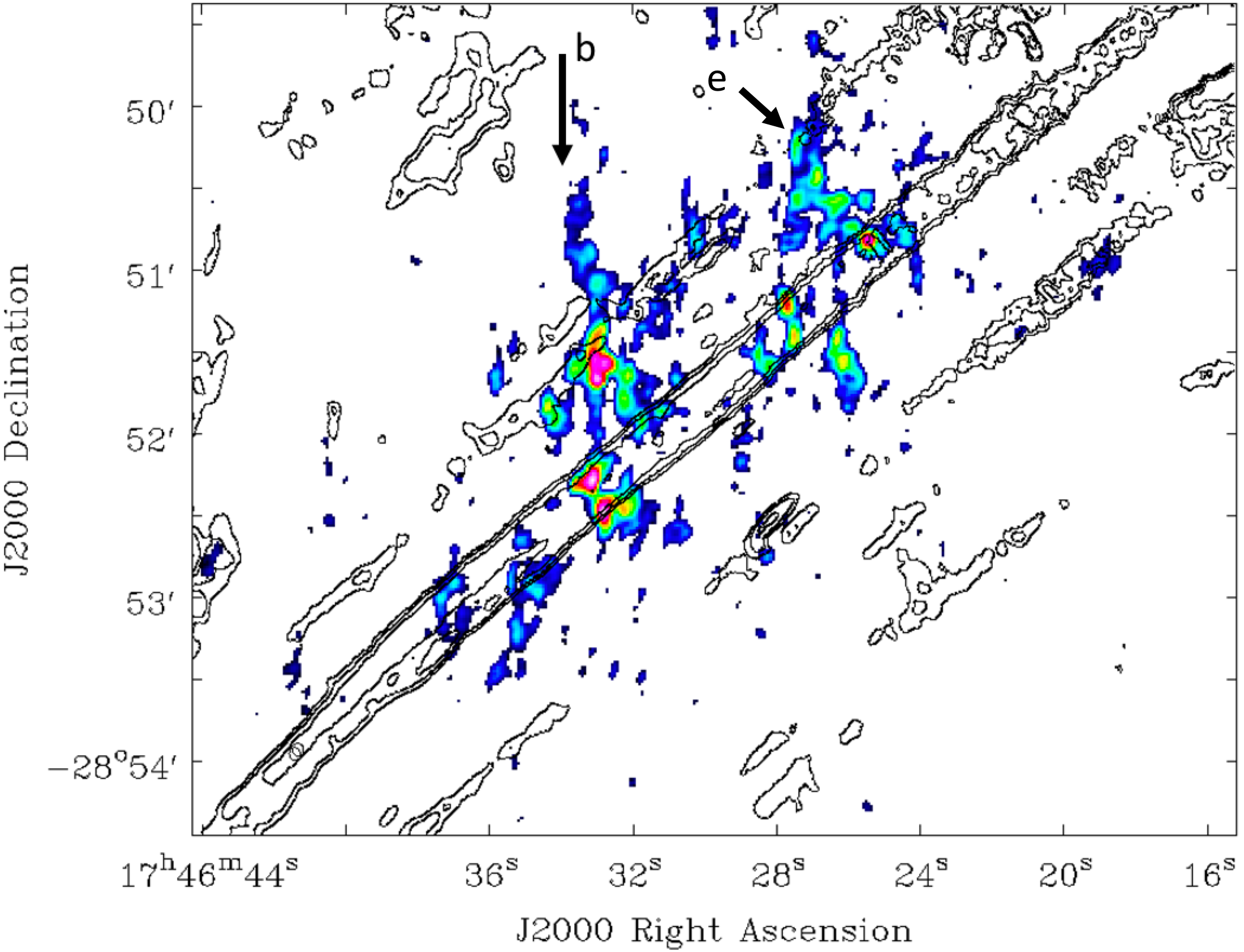}{0.7\textwidth}{}}
\caption{The colorscale shows the polarized intensity in the Arc NTFs at 10.4 GHz  with a resolution of 1.6\arcsec~x 0.9\arcsec~(upper panel) and at 4.9 GHz with a resolution of 6.5\arcsec~x 4.9\arcsec~(lower panel). The elongated structures discussed in Section \ref{sec:pol_elong} are named ``a''-``e'' and are marked with black arrows.} 
\label{fig:ps}

\end{figure*}

\begin{figure}[htp]

\plotone{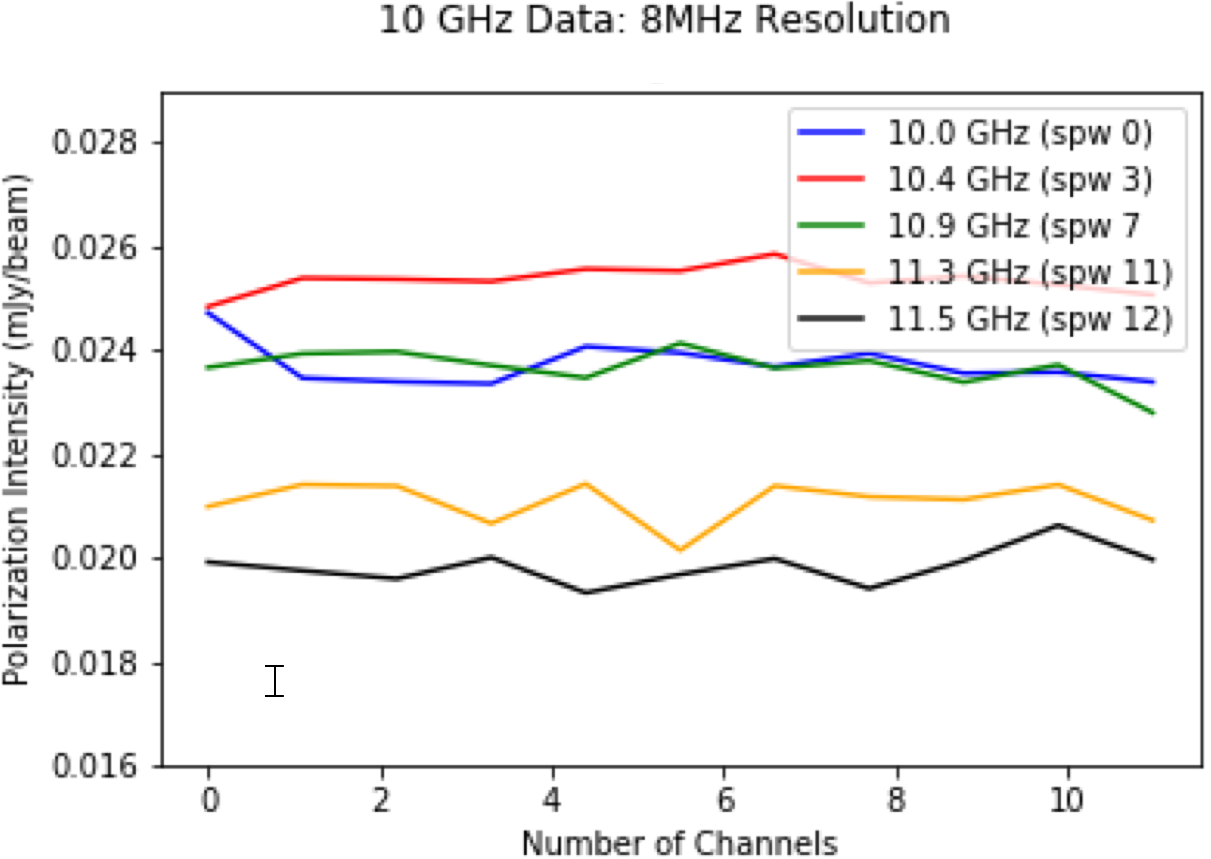}
\plotone{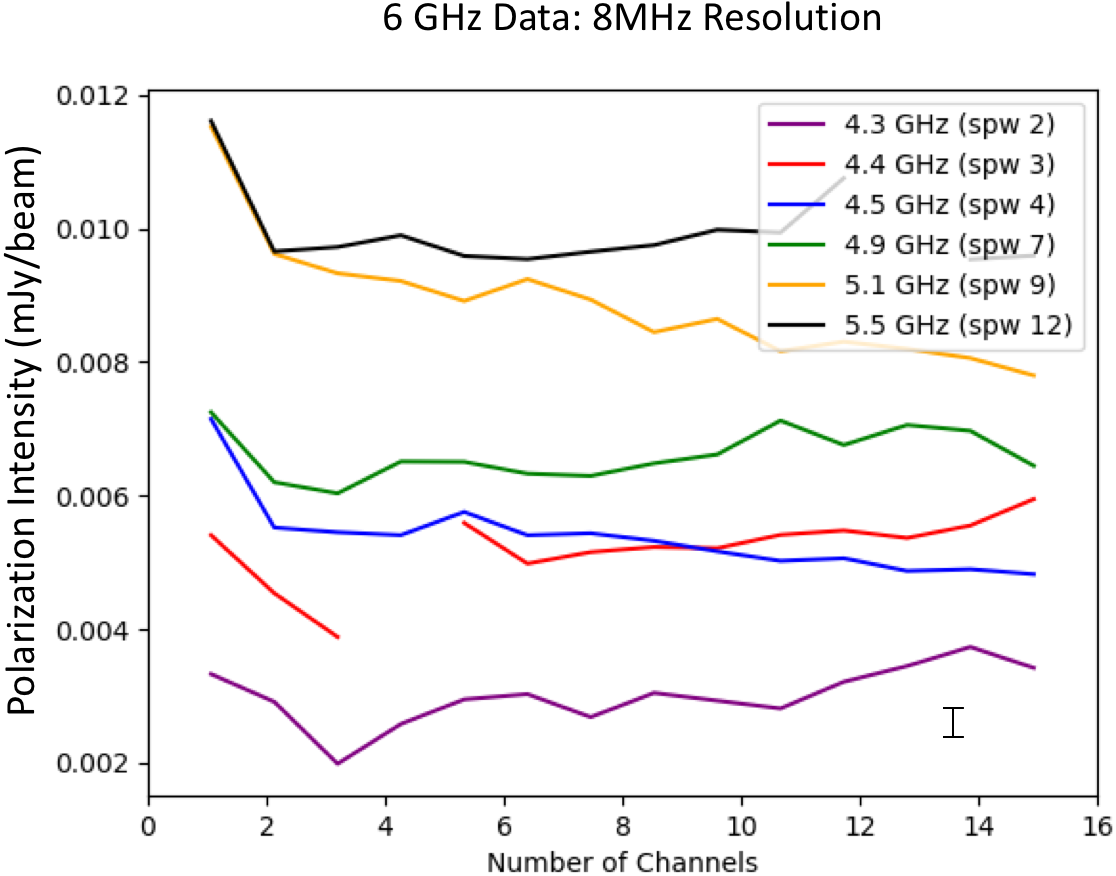}
\caption{Polarized intensity vs channel number for channels with a 8 MHz width. Data are taken from the 10 GHz (upper panel) and 5 GHz (lower panel) polarized intensity distributions and the pixel used to produce these plots is located at [RA,DEC]=[17h 46m 37.2s, - 28$\rm^{\circ}$ 53\arcmin{} 14\arcsec]. The error bars shown in the panels represent the characteristic uncertainties in the polarized intensity values. The legend indicates which spw and frequency is represented for each line.}
\label{fig:pchan}
	
\end{figure}

Figure \ref{fig:ps} shows the polarized intensity distribution at 10 GHz (top) and 5 GHz (bottom). The region shown in Figure \ref{fig:ps} is an inset from Figures \ref{fig:tot_I_2} and \ref{fig:tot_I_3}, showing only the portion of the figures where polarization is present. The polarized intensity is mostly confined within a 4\arcmin{} x 4\arcmin{} (9 pc by 9 pc). In both images, the polarization diminishes to the noise level before reaching either the southern end of the NTFs (south-eastern edge of the upper panel of Figure \ref{fig:ps}) or the northern region of the NTFs where the Sickle and Pistol are located.

The morphology of the polarized intensity differs from that of the total intensity. The polarized intensity appears to be organized into "clumps", which have been observed previously in observations of the NTFs in the GC region (e.g. \citet{YM1987,Morris1996}). Both panels in Figure \ref{fig:ps} show that regions of significant polarized intensity are separated by regions of complete depolarization. We took slices of the polarized intensity distribution to estimate the scale sizes of the clumping in the polarized intensity. We obtain characteristic angular sizes of $\rm\sim$10\arcsec~(0.4 pc) at 10 GHz and $\rm\sim$5\arcsec~(0.2 pc) at 5 GHz for the polarized intensity clumps. These values are within the same range of sizes that have been observed for other NTFs (\cite{YWP1997,Yusef-Zadeh2007,LaRosa2004,Lang1999b,Gray1995,Inoue1989}). 

%The clumpy appearance of the polarized intensity has been used to support the idea that there may be an irregular intervening medium which derotates the polarized emission in some regions more than others (see \citet{YMC1984, YM1987, Morris1996} and references therein). \citet{Gray1995} reaches a similar conclusion in their analysis of the Snake NTF. They note that the clumpiness of the polarized intensity could be due to a structure in an intervening medium that has density fluctuations which occur on scale sizes of the polarized intensity clumps. For our data that would imply the existence of an external medium which possesses spatial variations on sizes of 5 - 10\arcsec, which corresponds with physical sizes of 0.2$-$0.4 pc assuming the medium is local to the GC. The topic of rotating media toward the Arc NTFs will be explored in more detail in Section \ref{sec:rot_mech}.

We can identify four locations of concentrated polarized intensity along the central Arc NTFs at 10 GHz (bright red regions in Figure \ref{fig:ps}; top panel). Three regions of concentrated polarized intensity are present along the NTFs at 5 GHz (Figure \ref{fig:ps}; bottom panel). Peak polarized intensity values in these regions range from 4.9 to 6.5 mJy beam$\rm^{-1}$ at 5 GHz and 3.0 to 3.6 mJy beam$\rm^{-1}$ at 10 GHz. Calculating the percentage polarization is challenging for this source because the Arc NTFs are larger than the largest anglar size (LAS) we are sensitive to in our observations. The LAS of our 10 GHz observations ranges from 17 to 145\arcsec~while those of the 5 GHz data set ranges from 29 to 240\arcsec. While the polarized intensity is unlikely to be coherent on such large angular scales, we may lose some sensitivity to the largest total intensity features. This would make our percentage polarizations artificially higher. This can result in artificially high values for the percentage polarization. Despite this complication we can attempt to measure the total intensity and estimate a percentage polarization at locations of peak polarized intensity. We choose pixels that contained significant polarization at both 5 and 10 GHz. We find percentage polarization values of about 40\% at 10 GHz and about 10\% at 5 GHz, where the total intensity value used was drawn from the 3 GHz data set assuming the Arc NTFs have a flat spectral index \citep{Yusef-Zadeh1987a}. 

Though the polarized intensity is confined within the same region for both 10 and 5 GHz (refer to Figure \ref{fig:ps}), the polarization is far more sparse at 5 GHz than at 10 GHz. Clumps of polarization at 5 GHz occur on smaller sizes (average size of 5\arcsec{} or 0.2 pc) than those at 10 GHz (average size of 10\arcsec{} or 0.4 pc) and also more regions of depolarization between polarized clumps. There are larger gaps between polarized regions in the 5 GHz data set than is seen at 10 GHz.

\subsection{Polarization Along the Arc NTFs} \label{sec:ntf_pol}
The majority of the polarized intensity in both panels of Figure \ref{fig:ps} appears to be distributed along the length of the NTFs. At 10 GHz, the bright clumps of polarization are almost entirely organized along the central bundle of NTFs, as well as along the northern bundle of NTFs (see top panel in Figure \ref{fig:ps}). There is little polarized intensity associated with the southern bundle of NTFs. There is also significant polarization at 10 GHz located between the northern and central bundles of NTFs with an appearance of ``connecting'' these structures. There is not significant polarization between the central and southern bundles of NTFs, with only weak polarization located at several positions in the bundle of NTFs to the South of the central bright concentration. At 5 GHz, the features are similar, but much less extensive (Figure \ref{fig:ps}, bottom panel).

\subsection{Elongated Polarized Structures} \label{sec:pol_elong}

As described in Section \ref{sec:ntf_pol} there are features in the polarized intensity nearly orthogonal to the long axis of the Arc NTFs. These features correspond with the ``thorns'', previously identified by \citet{Inoue1989} in 15 GHz images of the polarized intensity of the Arc NTFs. In order to make it easier to identify these features, these vertical structures in polarization are marked with black arrows (a$-$e in Figure \ref{fig:ps} (upper)) and just ``b'' and ``e'' in Figure \ref{fig:ps} (lower panel)). These elongated features have no corresponding total intensity structures. The percentage polarizations of these structures are all much greater than 100\% because they lack total intensity counterparts. We note here that the lack of a total intensity counterpart is likely caused by the limits of our interferometric observations. If the elongated polarized structures are tracing a total intensity structure which is cohesive on scales larger than the LAS of our observations, the total intensity structure would not be detected.

These vertical structures are not as significantly represented in 5 GHz as in the 10 GHz image of polarized intensity. There are five clearly identified structures in the 10 GHz data, but there appears to be only two easily identifiable elongated structures in the 5 GHz polarized intensity image. These 5 GHz structures are spatially coincident with structures b and e in the 10 GHz distribution, indicating they are likely the same structures.

In both data sets the elongated polarized structures are confined within the physical extent of the NTF bundle. This is clearly discernible by comparing the total and polarized intensity distributions in Figure \ref{fig:ps}. We can see that for both 5 and 10 GHz data sets the majority of the polarized intensity is found within the boundaries traced by the northern and southern total intensity NTF bundles.

%Referring to the placement of the vertical structures throughout the 10 GHz data set we can see that they are evenly spaced throughout the field of view. Furthermore, these structure meet up with peak polarized intensity regions when crossing peak total intensity contours as is seen in both panels of Figure \ref{fig:ps}.

\subsection{Depolarization} \label{sec:freq_depol}

As described above, Figure \ref{fig:ps} illustrates that the polarized intensity is more sparse in the 5 GHz image than in the 10 GHz image. To investigate in more detail how the polarization varies across each of the observing bands, we can examine the data over the frequency range of our observations. As described in Section \ref{sec:meth}, observations were made across at least 10 spectral windows for each observing band, and these spectral windows each have a bandwidth of 128 MHz. Further, the spectral windows can be separated into numerous channels of selected bandwidths. 

Figure \ref{fig:pchan} shows the polarized intensity as a function of channel number for a representative position in each of the 10 and 5 GHz polarized intensity images. The pixel chosen to produce these plots is located in one of the regions in Figure \ref{fig:ps} that is bright at both 5 and 10 GHz. In the panels of Figure \ref{fig:pchan}, each line displays the polarized intensity as a function of channel number for different spectral windows within the observing bands at 10 and 5 GHz. These plots were made using channel widths of 8 MHz, therefore resulting in $\sim$16 channels for each spectral window. Sometimes at the edge of the spectral window, channels were removed due to poor response toward the edge of the spectral window. Similar plots were made for varying channel widths, including 2 MHz, 4 MHz, and 16 MHz; all channel widths resulted in similar plots so we only show the 8 MHz results here. 

We first investigate how the polarized intensity varies over spws 0-12.  The lowest frequency spw in the upper panel of Figure \ref{fig:pchan} contains the lowest average polarized intensity. As frequency increases, the average polarized intensity of the corresponding spw increases. Conversely, in the lower panel of Figure \ref{fig:pchan}, there is no such frequency-dependent trend in the polarized intensity. Indeed, throughout this 10 GHz data set the polarized intensity is confined between 20 - 26 mJy beam$\rm^{-1}$.

We can next investigate how the polarized intensity varies over channel number for each of the spws. In the upper panel of Figure \ref{fig:pchan} the spws maintain a generally constant polarized intensity level over channel number except for the 5.1 GHz spw (spw 9). This one demonstrates a clear decrease in polarized intensity as a function of channel number. The lower panel of Figure \ref{fig:pchan}, conversely, shows no such variation with the spw. All the spws maintain the same average polarized intensity value with some systematic variation as a function of channel number. 

%Comparing the polarized intensity values obtained in the upper and lower panels of Figure \ref{fig:pchan}, we see that the 10 GHz polarized intensity ranges from 20 to 26 mJy beam$\rm^{-1}$ whereas the 6 GHz data ranges from 18 to 9 mJy beam$\rm^{-1}$. This indicates that the 10 GHz polarized intensity values are systematically higher than those at 6 GHz. 
%(\textbf{DYLAN-was this work done with 7" beams or "full" resolution - i.e., different resolutions for the 6 and 10 GHz - Hailey's webpages might describe}).
	
\subsection{Rotation Measure Results} \label{sec:mag_meas}
We developed a script to perform a linear fit of the $\rm\chi$ vs. $\rm\lambda^2$ data generated by our observations. This package utilized the Python LMFIT package, which implements a regressive least-squares fitting technique to a data set using a user-specified model \citep{Newville2016}. The rotation measure (RM [$\rm rad\,\,m^{-2}$]) and intrinsic polarized intensity angle ($\rm\chi_0$ [rad]) of a line of sight are defined as the slope and y-intercept of the best fit line, respectively:
\begin{equation}
\rm \chi = RM\lambda^2 + \chi_0 \label{eq:LS}
\end{equation}
where $\rm\lambda^2$ [$\rm{}m^2$] is the wavelength squared and $\rm\chi$ [rad] is the polarized intensity angle detected by the observer.

The RM is an integral along the line of sight of the product of the electron density and line of sight magnetic field strength:
\begin{equation}
    \rm RM = \frac{e^3}{2\pi{}m_e^2c^4}\int_0^L\!n_eB_{||}\,dl \label{eq:RM}
\end{equation}
where e is the charge of an electron, $\rm{}m_e$ is the mass of an electron, c is the speed of light, $\rm{}n_e$ is the electron number density, $\rm{}B_{||}$ is the strength of the magnetic field along the line of sight, L is the distance between the source and the observer, and dl is the deferential integration element along the line of sight.

\begin{figure*}[htp]

\gridline{\fig{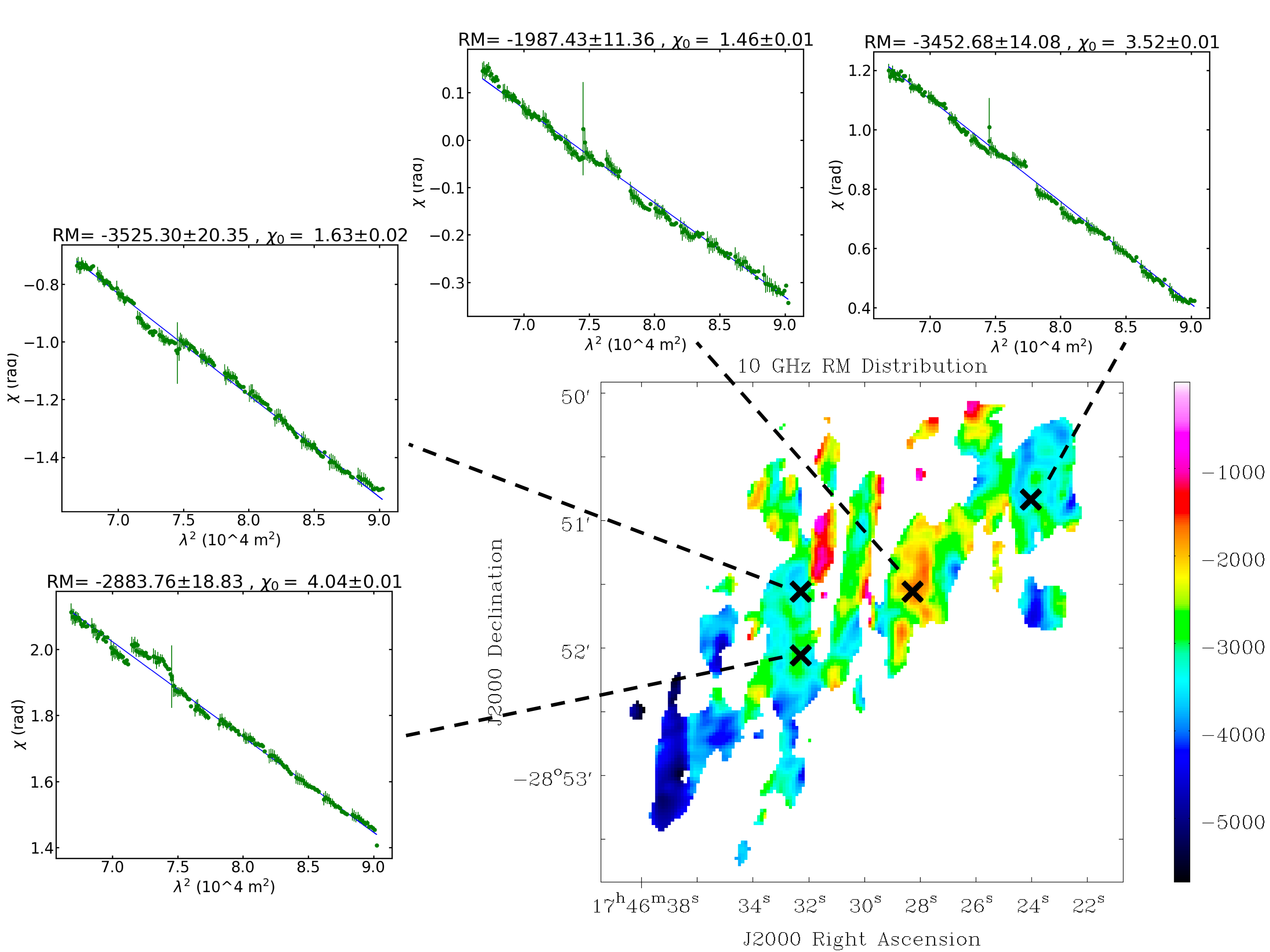}{0.9\textwidth}{}}
\caption{The 10 GHz RM distribution with four examples of the least-squares method used to determine the RM values. The color scale of the RM distribution details the range of RM values, measured in rad m$\rm^{-2}$. The pixels used to demonstrate the quality of the least-squares fits are marked on the RM distribution with black `X' symbols. The green data points in the fitting plots represent values of the polarization angle as a function of $\rm\lambda^2$ and the blue lines show the best fit linear models to the data.}
\label{fig:RM_supa}

\end{figure*}

Figure \ref{fig:RM_supa} shows the distribution of RM obtained at 10 GHz from fitting the polarization angle as a function of $\rm\lambda^2$ across the field of view. The insets in Figure \ref{fig:RM_supa} show plots of polarization angle as a function of $\rm\lambda^2$ for four representative pixels. These plots illustrate the larger number of measurements ($\rm\sim$200) that are used to generate the fits and the quality of the least-squares fits to the polarization angle data. The pixels chosen exhibit fairly high polarized intensity and were also chosen to illustrate a range of RM values. 

The values of RM fit to the data and shown in Figure \ref{fig:RM_supa} range from $-$500 to $-$5500 rad m$\rm^{-2}$. Most measurements have magnitudes generally greater than 1000 rad m$\rm^{-2}$. We can see from the RM distribution that the eastern and western regions of the map display RMs of larger magnitude generally in the range of 3500$-$5500 rad m$\rm^{-2}$. In the center of the distribution the RM magnitudes become generally lower with magnitudes ranging from 1500$-$2500 rad m$\rm^{-2}$.

The examples of the least-squares fitting routine indicate that the polarization angle does follow a linear trend as a function of $\rm\lambda^2$. There is a channel that contains substantially larger error bars, as can be seen in all of the examples shown in the figure. This pixel corresponds to an edge channel of an spw where the telescope sensitivity is reduced. Therefore, the signal-to-noise ratio in this channel is particularly low, explaining its large error and deviation from the general trend of the rest of the data. Other edge channels have been removed from this data set, with this one retained as an example. It is important to note, however, that because of the large numbers of channels used to generate these fits ($\rm\sim$200), deviations for individual data points from the rest of the data have negligible effects on the qualities of the best-fit models.

We also estimate the errors on our RM values by analyzing output fitting errors produced from the LMFIT covariant matrix for each line of sight. Our errors on RM range from 10$-$130 rad m$\rm^{-2}$, an order of magnitude below the RM values or smaller. The RM errors are largest at the spatial edge of the RM distribution where the polarized intensity is the weakest.

\subsection{Intrinsic Magnetic Field Distribution} \label{sec:mag}

\begin{figure*}[htp]

\gridline{\fig{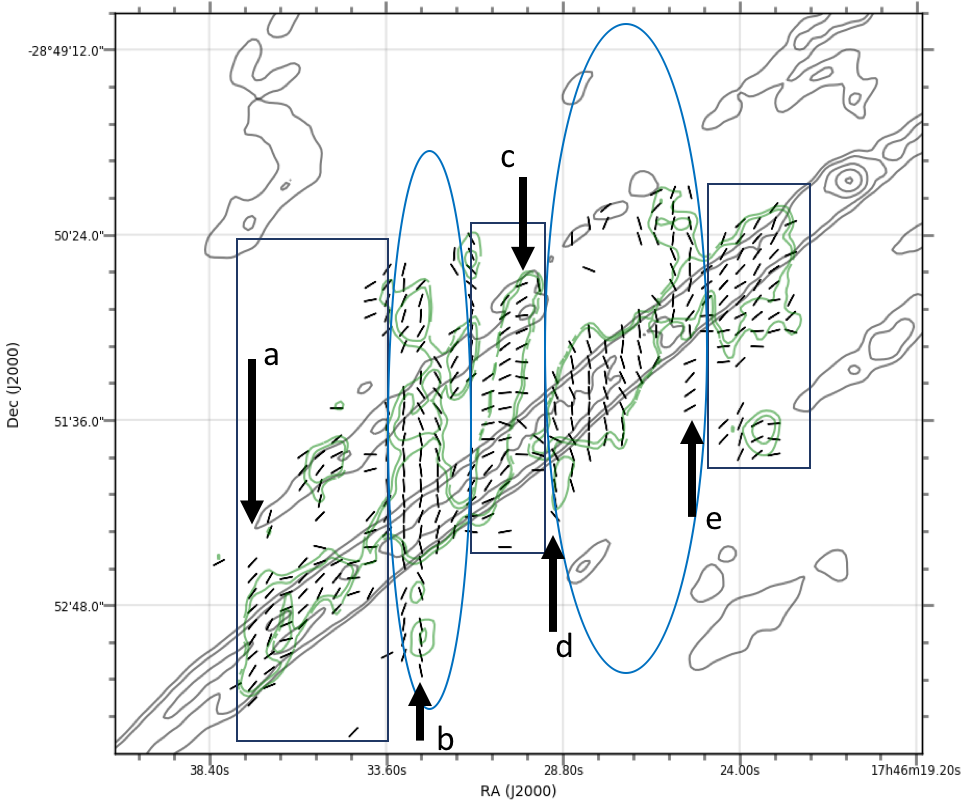}{1.0\textwidth}{}}
\caption{Vectors represent the intrinsic magnetic field orientation in the Arc NTFs as determined from the 10 GHz data. These vectors are overlaid on contours of 10 GHz total intensity (black) and polarized intensity (green). Contours of total intensity are shown at levels of 1, 2, 4, 8, 16, and 32 times the rms noise level of 2.43 mJy beam$\rm^{-1}$ and polarized intensity at levels of 0.2 and 0.3 times the maximum polarized intensity value of 23.2 mJy beam$\rm^{-1}$. Elongated polarized structures ``a'' through ``e'' are labeled. The elliptical regions show areas where the magnetic field is predominantly at a 60 degree angle with respect to the Arc NTFs and the rectangular regions show areas where the magnetic field is predominantly parallel to the Arc NTFs.}
\label{fig:vec_over_x}

\end{figure*}

Using the rotation angles as a function of wavelength-squared we derive in Section \ref{sec:mag_meas}, we can de-rotate the measured electric field orientations to the intrinsic orientation in the source using Equation \ref{eq:LS}. These intrinsic polarization vectors are then rotated by an additional 90 degrees to convert from intrinsic electric field orientations to source magnetic field orientations. Due to the better signal-to-noise of the polarization at 10 GHz, we present intrinsic magnetic field results based on the 10 GHz data. Figure \ref{fig:vec_over_x} displays the 10 GHz intrinsic magnetic field distribution resulting from the application of this technique.

Figure \ref{fig:vec_over_x} shows that the distribution of the intrinsic magnetic field across the Arc NTFs is complex. One of the notable features is that there are multiple parallel NTFs where the magnetic field orientation can be determined (i.e., the northern bundle of NTFs as well as the central bundle of NTFs). Most previous studies have determined the magnetic field in NTFs that are isolated and single-stranded, with the exception of the Pelican \citep{Lang1999a}, for which the intrinsic magnetic field is determined for substructure in a system of several NTFs. Although the orientation of the magnetic field varies across the region studied, a pattern of fairly well-ordered behavior is present and appears to alternate across the Arc NTFs. There are regions where the magnetic field lines are predominantly parallel to the Arc NTFs  (indicated by rectangular boxes) and regions where they are oriented at a 60 degree angle to the long axis of the the Arc NTFs (elliptical regions). For reference, the elongated polarized structures described earlier are labeled ``a'' through ``e.''

We examine the detailed structure of the intrinsic magnetic field in this section, describing the features present in Figure \ref{fig:vec_over_x},  from east to west. In the region at the furthest eastern extent, the orientation of the magnetic field is primarily organized parallel to the NTFs. There are two zones in this region where polarization (and hence magnetic field) is measured $-$ along the central bundle of NTFs and also a zone along a bundle of filaments to the north. Both zones show similar structure in the magnetic field. For the adjacent region the magnetic field vectors almost immediately change orientation by about 60 degrees. The distribution of polarized intensity in this region (shown in the ellipse) is elongated (feature ``b'') and extends between the NTF bundles, with a faint polarized clump (and oblique magnetic field vectors) appearing to be associated with the Southern bundle of NTFs. Moving to the West in Figure \ref{fig:vec_over_x}, the magnetic field then changes back to being oriented along the NTFs in the next region of bright polarized intensity, the region containing structure ``c,'' which extends continuously between the NTF bundles. Along the NTFs near RA=17h 46m 29s, DEC=$-$28d 51\arcmin~42\arcsec, there is a region where it is obvious that the magnetic field vectors are changing from a 60 degree orientation to a parallel one. The vectors rotate through about 75 degrees in this small region indicating that some mechanism is causing a strong rotation. 

The region indicated by the ellipse (also located between polarization structure ``d'' and ``e'') is strongly polarized and primarily  associated with the central bundle of Arc NTFs. The magnetic field orientation in this region is aligned along the NTFs with slight variations along the edges of the clumps. There is some weak polarization associated with the northern bundle of NTFs; in this region, there are not many vectors and they appear to be oriented along the northern NTFs. 

In the furthest western extent of Figure \ref{fig:vec_over_x}, two clumps of polarized emission are apparent, associated with the central bundle of NTFs as well as the southern NTFs. Both regions show vectors primarily organized along the length of the NTFs with some variation of rotation of the vectors along the edges of the clumps of polarization, where presumably the signal to noise in the polarized intensity drops. It is also notable that all 5 of the elongated structures in polarization (regions ``a'' through ``e'') occur near regions where the magnetic field orientation changes dramatically between rotated and parallel orientations.  In all 5 regions 80\% or more of the orientations are aligned with one another.

\section{DISCUSSION} \label{sec:disc}

\subsection{Large-scale Total Intensity Structure} \label{sec:tot_I_disc}
As discussed earlier (Section \ref{Sec:I_struct}), our total intensity images show further evidence for a large-scale and possibly helical structure that surrounds the Arc NTFs. This large-scale structure is characterized by the diffuse but prominent helical segments we mark in Figure \ref{fig:tot_I_1}. There have been several ideas for what this diffuse curved structure could be. \citet{Benford1988} postulates that this structure could be produced through a magnetic force-free configuration characterized by multiple magnetic field pinches along the length of the Arc NTFs. There is also evidence that the Arc NTFs are located in a region that is strongly affected by the ionization from two very massive stellar clusters. On scales similar to the diffuse helical structure, and overlapping somewhat with the helical segments is a large radio and infrared shell. This radio shell is observed at infrared wavelengths from 10 - 38 $\rm\mu$m and is centered approximately on the center of our field of view and encompasses the Arc NTFs, the Sickle HII region and Quintuplet stellar cluster. The Quintuplet Cluster is located roughly at the center of the 10\arcmin~(23 pc) wide Radio Bubble, and it is possible that the Radio Bubble is a shock front of winds produced by this cluster \citep{Simpson2007}.

\subsection{Elongated Polarized Structures ("Thorns")} \label{sec:thorn_disc}

Our data also provide evidence for elongated polarized structures consistent with those originally presented by \citet{Inoue1989} (Section \ref{sec:pol_elong}). \citet{Inoue1989} identify two ``thorns'' in their polarized intensity images of the Arc NTFs at 15 GHz. The elongated features they detect correspond to the elongated structures ``b'' and ``c'' in our observations (Figure \ref{fig:ps}). Our results expand on the \citet{Inoue1989} results by revealing additional elongated polarized structures. 

As discussed in \citet{Inoue1989}, the elongated structures could either be local to the Arc NTFs and embedded in the NTF system or foreground magnetized structures. We saw in Section \ref{sec:pol_elong} that while the elongated structures are generally perpendicular to the extent of the Arc NTFs their polarized emission is confined within the extent of the NTF bundle (refer to Figure \ref{fig:ps}). If the elongated structures were foreground objects unassociated with the Arc NTFs, we would not expect the extent of the polarized intensity to be dependent on the extent of the total intensity of the Arc NTFs.

The elongated structures could be local to the Arc NTFs because of the relationship between the extent of the polarized intensity and the NTF bundle. \citet{Inoue1989} theorize that if the elongated structures are local to the NTFs they may be the source of relativistic electrons producing the synchrotron emission seen throughout the Arc NTFs. The mechanism for this acceleration could be the interaction of the NTF magnetic field system with that of the elongated structures, causing magnetic reconnection at the locations where the fields intersect.

An alternative explanation which we conder is that the elongated structures are not tracing a separate magnetic field from the Arc NTFs, but that the polarized intensity which extends vertically away from the NTFs is ubiquitous throughout the Arc NTF length. The reason for the unpolarized gaps between the elongated structures in this scenario would be depolarization caused by intervening Faraday screens. Depolarization becomes more prevalent at larger wavelengths, so we can compare our 10 and 5 GHz polarized intensities to determine whether the elongated structures are by-products of depolarization or not.

While 5 separate elongated structures are identified in the 10 GHz data set, only two are observed in the 5 GHz data set (``b'' and ``e'')). Since none of the other structures are observed at 5 GHz, it is possible that they have been depolarized at the lower frequencies of the 5 GHz data set. The larger degrees of depolarization experienced towards these other elongated structures could indicate the presence of higher density intervening media at the locations of these structures. 

A final possibility for the location of the thorns with respect to the Arc NTFs is that they are slightly foreground to the Arc NTFs. These structures are similar in appearance with polarized threads found in the Galactic disk (\citet{Jelic2018} and references therein). As with our elongated polarized structures, these Galactic disk features have no significant total intensity counterparts and as a result have fractional polarizations far in excess of 1.0 \citep{Wieringa1993}. These disk structures are much larger than the ones seen towards the Arc NTFs, occurring on angular sizes of degrees, \citet{Jelic2015,Jelic2018}). The polarized filaments are separated by depolarization canals which \citet{Jelic2015} theorize are caused by beam depolarization. Because of their Faraday depth properties, these polarized structures are thought to be part of a larger interconnected structure which is highly foreground with respect to the background polarized emission \citep{Jelic2015}.

A similar physical situation could be occurring in our elongated structures. The regular spacing of the elongated structures could indicate regularly-spaced gaps in a foreground rotating screen. The elongated structures shown in Figure \ref{fig:ps} could therefore be a component of the possible helical structure identified by \citet{YM1987}. This idea will be further discussed in Section \ref{sec:helix}.

\subsection{Frequency Dependent Depolarization} \label{sec:depol_disc}
\begin{figure}
    \gridline{\fig{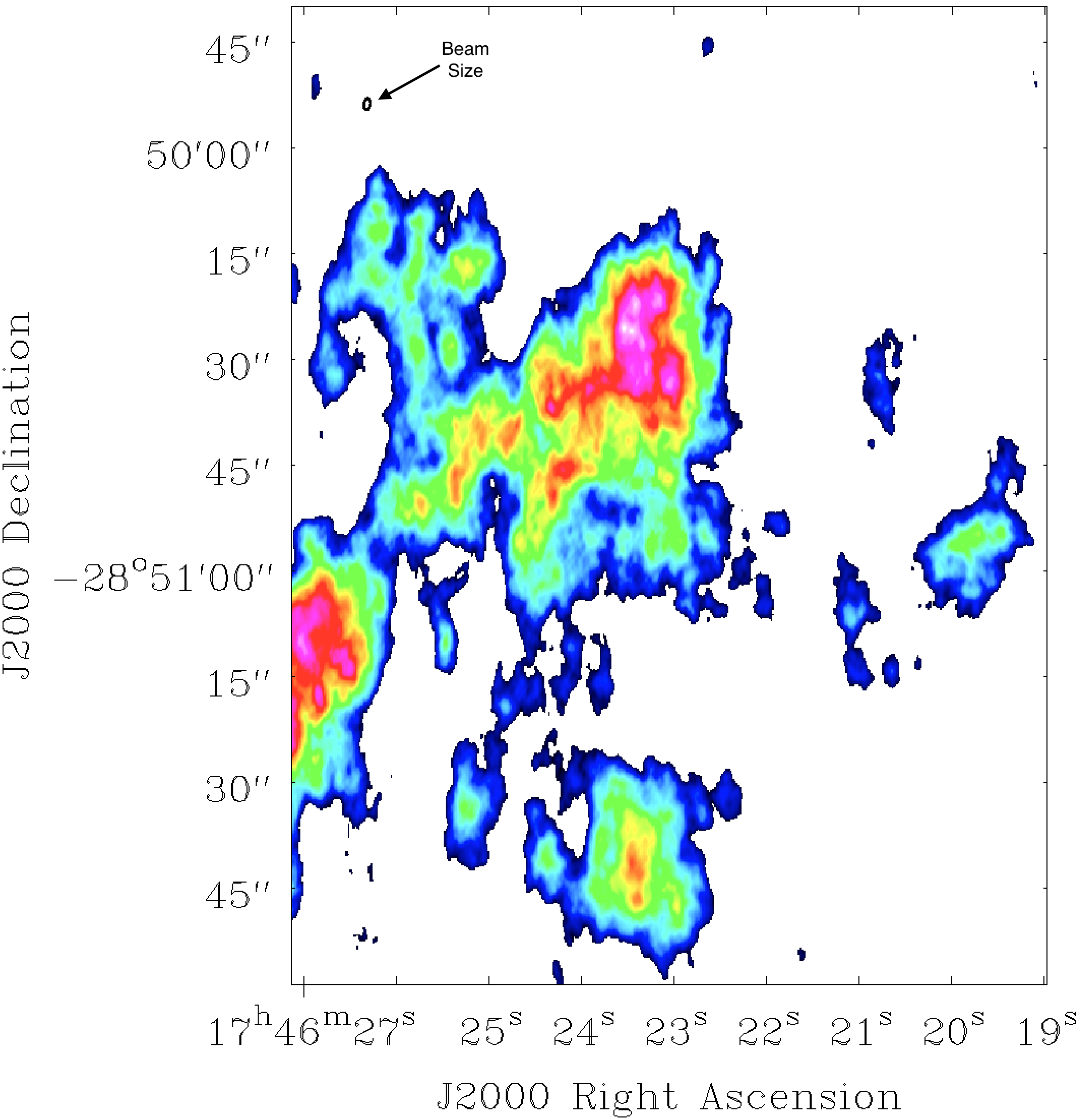}{0.4\textwidth}{}}
    \gridline{\fig{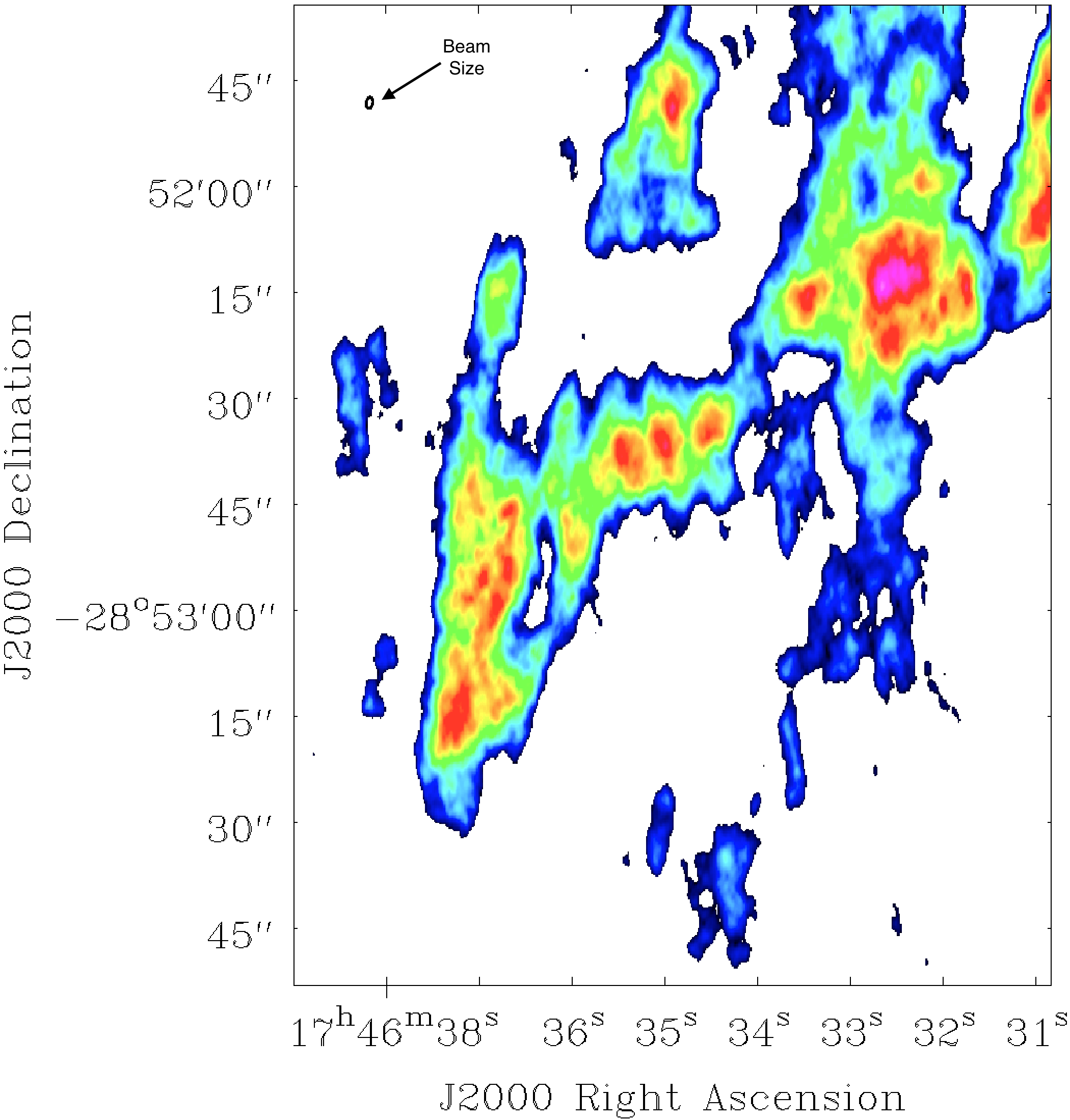}{0.4\textwidth}{}}
    \caption{Two zoomed in regions of the 10 GHz polarized intensity distribution from Figure \ref{fig:ps}. The beam size is shown and marked in the upper left corner of the panels.}
    \label{fig:p_beam_compare}
\end{figure}

%\begin{figure*}[tp]

%\gridline{\fig{rm_w_p.png}{0.9\textwidth}{}}
%\caption{RM distribution of the Arc NTFs with contours of the smoothed polarized intensity at 0.2, 0.4, 0.6, 0.8 and 0.9 times the peak value of 23.2 mJy beam$\rm^{-1}$. Both distributions are drawn from the 10 GHz data set. Values for the color bar are in units of rad m$\rm^{-2}$.}
%\label{fig:rm_p_compare}

%\end{figure*}

In Section \ref{sec:freq_depol} we outline the observed frequency dependent depolarization. We see in the panels of Figure \ref{fig:pchan} that the polarized intensity at 5 GHz is systematically lower than at 10 GHz. Furthermore, in Figure \ref{fig:ps} we can see that the polarized intensity distribution is far more extended in the 10 GHz contours than in the 5 GHz color scale. We would like to assess what underlying mechanisms are causing this frequency dependent depolarization. 

\subsubsection{Beam Depolarization} 
There are a number of mechanisms which could cause the frequency dependent depolarization we observe. The first mechanism we consider is beam depolarization. Beam depolarization occurs when different polarization angles are averaged over the resolving beam of the telescope, or when the magnetic field varies in strength and/or orientation within the beam \citep{Burn1966}. The depolarization could originate from the emission source itself or from an intervening medium along the line of sight.

To evaluate whether beam depolarization is a significant effect on our data we can compare the size of the restoring beam with the size scales of the polarized intensity clumps. In Figure \ref{fig:p_beam_compare} we see that the polarized intensity clumps are generally much larger than the 1.6\arcsec~x 0.9\arcsec~restoring beam size. For example, the top panel of Figure \ref{fig:p_beam_compare} reveals polarized intensity structure that is coherent to tens of arcsecond scales, $\rm\geq$ 10 times the beam size. The bottom panel, while not as extreme, does show polarized intensity clumps that are $\rm\geq$ 9\arcsec~on a side. We would not expect to detect clumps of polarized intensity on sizes larger than our restoring beam size if beam depolarization were a significant effect. 

We also smoothed the polarized intensity distributions at 10 and 5 GHz by doubling the restoring beam sizes. This allowed us to make an additional assessment of the importance of this mechanism. If beam depolarization is a significant effect, smoothing the data in this manner should reduce the polarized intensity values we obtain. At both 5 and 10 GHz the doubled resolving beam size does not reduce the polarized intensity values. Furthermore, the polarized intensity distribution does not drastically change in morphology with such a smoothing. As a result of these tests, we assess beam depolarization as insignificant in our data set.

\subsubsection{Bandwidth depolarization}
Bandwidth depolarization occurs when different polarization angles are detected over a given frequency width ($\rm\Delta\nu$), where $\rm\Delta\nu$ is the frequency resolution of the observations. We have seen in Section \ref{sec:freq_depol} and the panels of Figure \ref{fig:pchan} that at $\rm\Delta\nu$ widths of 4 - 16 MHz no significant change in polarized intensity is observed between different channels for most of the spws in our data sets. If the polarized intensity were to change as a function of frequency, it would indicate a frequency dependence on the RM value, which in turn would indicate a frequency dependence on the polarization angle. Since the polarized intensity is fairly constant, it would imply no significant changes in polarization angle. However, at the lowest frequency range of the 5 GHz data set, the polarized intensity increases with increasing channel number as can be seen with the 4.5 GHz spw in Figure \ref{fig:pchan}. Although, because only this low frequency range of our data set demonstrates this feature, it does not explain the depolarization we observe across our two data sets. We therefore rule out bandwidth depolarization as a significant effect in our data set as a whole. It therefore remains unclear which mechanism is responsible for the depolarization we observe.

\subsection{RM Analysis of the Radio Arc} \label{sec:RM_disc}

\subsubsection{RM Magnitudes and Direction} \label{sec:RM_mags_disc}

As discussed in Section \ref{sec:mag_meas}, we obtain RM values which range from -500 to -5500 rad m$\rm^{-2}$ (Figure \ref{fig:RM_supa}). This range of RM values agrees with previous measurements of the Arc NTFs, which find RM values of about -1660 and -5500 rad m$\rm^{-2}$ for the Arc NTFs \citep{Tsuboi1986,Yusef-Zadeh1986a,YM1987}.

We can assess how the range of RM magnitudes from the Arc NTFs compares with those observed from the rest of the NTF population. \citet{Lang1999b} find RM magnitudes for the northern Thread at 6 GHz which vary from 100 to 2300 rad m$\rm^{-2}$. \citet{YWP1997} find RM magnitudes ranging from 370 to 4200 rad m$\rm^{-2}$ for G359.54+0.18 at 6 GHz, with the bulk of the RM magnitudes being greater than 1000 rad m$\rm^{-2}$. \citet{Gray1995} find RM magnitudes ranging from 2000 - 5500 rad m$\rm^{-2}$ for G359.1 - 00.2 at 6 GHz. These NTFs possess RM magnitudes comparable to our results for the Arc NTFs. Conversely, \citet{Lang1999a} finds a less variant RM distribution ranging from only 0 to 1000 rad m$\rm^{-2}$ in magnitude for the Pelican at 6 GHz. 

\citet{Law2011} compare the RMs of various NTFs with the RM magnitudes obtained from polarized continuum emission within a 0.5 deg$\rm^2$ region of the GC. They find RMs of generally constant magnitude throughout their region of study which agrees with the individual observations of NTFs that find analogous RM magnitude ranges. This does not explain, however, why the Pelican of \citet{Lang1999a} possesses a more compact RM distribution than the other filaments. However the Pelican is an unusual NTF located over a degree away from Sgr $\rm A^*$ that is oriented parallel, rather than perpendicular, to the galactic plane. As such, it is not covered by the \citet{Law2011} region of study. Its unique location could explain its unique RM distribution.

Now we must explore how the RM values we obtain from the Arc NTFs compare with other filaments. While the Arc NTFs possess negative RM values, G0.08+0.15 and G359.1 - 00.2 are found to possess strictly positive RM values \citep{Lang1999b,Gray1995}. Conversely, the Pelican contains an equal mixture of positive and negative RMs whereas G359.54+0.18 contains entirely negative RM values \citep{Lang1999a,YWP1997}. \citet{Law2011} determine a four-quadrant mapping of the RM sign in the central region of the GC in which the filaments discussed above are found to agree in sign with the radio continuum measurements. However, while the NTF population traces the large-scale RM structure mapped by \citet{Law2011}, the Arc NTFs do not: possessing negative RM values in a quadrant that is otherwise dominated by positive RMs (refer to their Figure 14).

To assess whether the RMs obtained from the NTF population are reasonable, we can compare with the RMs obtained from other Galactic sources. H\textsc{ii} regions have RMs ranging from 100 to $\rm\approx$ 1000 rad m$\rm^{-2}$ \citep{Costa2016,Costa2018}. The rotation measures of pulsars local to the Milky Way (0.1 - 60 kpc away) reveal an extensive range of RM magnitudes from 10 to a few times 1000 rad m$\rm^{-2}$ \citep{Han2018}, though most of these pulsars have magnitudes of $\rm\approx$ 100 rad m$\rm^{-2}$. Therefore, while the RM magnitudes obtained from the NTF population are high, they are analogous to the RMs obtained from some HII regions and pulsars.

Other sources of high RM magnitudes have been found local to the GC. Sgr A$\rm^*$ was observed to have an RM magnitude of 43000 rad m$\rm^{-2}$, an order of magnitude larger than the RM magnitudes obtained from the NTF population \citep{Bower2003}. Furthermore, the GC magnetar PSR J1745-2900 possesses an RM magnitude of 66960 rad m$\rm^{-2}$, same order of magnitude as that of Sgr A$\rm^*$ (\citet{Desvignes2018} and references therein). Therefore it can be seen that while the RM magnitudes obtained from the Arc NTFs are large, they are within the range of previous observations and are an order of magnitude below the most extreme sources of RM seen in the GC.

\subsubsection{Assessing Possible Faraday Rotation Mechanisms} \label{sec:rot_mech}
There are three mechanisms that could be causing the distribution of RM values we observe:

\noindent\textbf{Differential Faraday Rotation (Depth depolarization):} Synchrotron emission produced from the far side of a source that is both emitting and rotating will be rotated by a different amount than the emission from the near side of the source \citep{Sokoloff1998}. This is a source of Faraday rotation of the polarization angle, but will also cause depolarization of the emission when averaging over the line of sight. Assuming the source is an emitting and rotating medium of uniform thickness and particle density in the presence of a uniform magnetic field, this mechanism is characterized by:
\begin{equation}
\rm P = p_0\frac{\sin\left(2RM\lambda^2\right)}{2RM\lambda^2}e^{2i(\chi_0 + RM\lambda^2)} \label{eq:DFR}    
\end{equation}
where P is the observed polarized intensity [Jy beam$\rm^{-1}$], $\rm{}p_0$ is the intrinsic polarized intensity emitted by the source [Jy beam$\rm^{-1}$], RM is the rotation measure of the rotating source [rad m$\rm^{-2}$], $\rm\lambda^2$ is the wavelength squared of the observation [m$\rm^2$], and $\rm\chi_0$ is the intrinsic polarization angle of the emission emitted by the source [rad].

This equation only holds while the assumption of a uniform magnetic field is valid. In situations where the magnetic field is non-uniform, the next mechanism is used to characterize internal rotation.

\noindent\textbf{Internal Faraday Dispersion:} This is a physical situation identical to depth depolarization where the source of polarized emission also causes rotation, but where the magnetic field within the source is non-uniform. It is characterized by the following equation:
\begin{equation}
    \rm P = p_0e^{2i\chi_0}\frac{1 -e^{4iRM\lambda^2-2\sigma_{RM}^2\lambda^4}}{2\sigma_{RM}^2\lambda^4-4iRM\lambda^2} \label{eq:IFD}
\end{equation}
where we assume a Gaussian distribution in RM across the source with a standard deviation of $\rm \sigma_{RM}$ [rad m$\rm^{-2}$]. This mechanism also causes both rotation and depolarization.

\noindent\textbf{External Faraday Rotation:} The Faraday rotation could also be caused by an intervening Faraday screen with a uniform magnetic field located along the line of sight \citep{Burn1966}. For purely foreground screens the equation for this mechanism is characterized by:
\begin{equation}
    \rm P = p_0e^{2i(\chi_0 + RM\lambda^2)} \label{eq:EFR}.
\end{equation}
This mechanism causes rotation of the emission, but will not depolarize it.

Because the magnetic field of the NTFs is parallel to the extent of the NTFs, there would be little to no magnetic field component along the line of sight. Since the RM is proportional to the strength of the line of sight magnetic field strength (Equation \ref{eq:RM}) there is little internal rotation in the Arc NTFs from our angle of observation. Conversely, the nature of the magnetic field in the elongated structures (if they are external sources) is less clear and so these structures could have a substantial line of sight magnetic field. As a result, we do not expect internal Faraday rotation to be a significant effect in the Arc NTFs, although it could be a factor in the elongated structures. As a result, we consider external Faraday rotation to be the main mechanism responsible for the rotation we observe.

\subsubsection{Predictions for Possible QU-Fitting Models}
Robust algorithms, like RM-Synthesis and QU-fitting can determine the presence of multiple RM components along the line of sight (e.g: \citet{Brentjens2005,OSullivan2012}). Our data is a strong candidate for the application of such algorithms given the complexity of the lines of sight towards the GC. Some models that could be applied to our lines of sight using QU-fitting based on the findings from Section \ref{sec:rot_mech} are: \textbf{1)} A model characterized by a varying number of external components each modelled by Equation \ref{eq:EFR} and \textbf{2)} A model identical to the first but with the elongated polarized structures possessing some internal Faraday rotation characterized by either Equations \ref{eq:DFR} or \ref{eq:IFD}.

\subsubsection{Nature of the ISM Towards the GC} \label{sec:nature}
In Section \ref{sec:rot_mech} we found that external Faraday rotation is significant in our data set. However, we cannot determine the number of rotating components along the line of sight due to the limitations of the least-squares analysis. We can review previous investigations into the GC ISM to gain some preliminary knowledge of the Faraday screen morphology.

\citet{Lazio1998a} searched for extragalactic sources in the GC region. They found a deficit of sources near Sgr A$\rm^*$ and found evidence of enhanced scattering diameters for the sources studied. Both of these findings indicate that there is a scattering medium local to the GC that possesses an angular size of $\rm\approx$ 0.5 square degrees \citep{Lazio1998b}.  They propose that the scattering medium has a high electron density ($\rm n_e =$ 10 $\rm cm^{-3}$) and consists of the interaction between molecular and X-ray gas in the GC. Its proximity to the GC implies that this scattering region also possessing a strong magnetic field. The combination of the high electron density and magnetic field strength indicates this medium could be a significant Faraday screen.

Recent observations of the Central Molecular Zone (central 200 pc of the GC) indicate a distribution of dense molecular gas and hot X-ray gas throughout the GC \citep{Simpson2018,Terrier2018,Kruijssen2015,Longmore2013}. Furthermore, observations of the Radio Arc complex itself have found evidence for molecular and X-ray gas local to the Arc NTFs (\citet{Butterfield2018,Yusef-Zadeh2007} and references therein). There is therefore ample observational evidence to indicate that this hyperstrong scattering region of \citet{Lazio1998b} exists along the line of sight to our observations of the Arc NTFs and could be a strong rotating medium. 

\subsection{Analysis of the Intrinsic Magnetic Field} \label{sec:mag_disc}

Previous RM observations of the Arc NTFs were made by fitting only a small number of rotation angles across the frequency band, so producing a reliable image of intrinsic magnetic field orientations was difficult in the work of \citet{YM1987}. However, they present the distribution of observed electric field orientation in the Arc NTFs. They find that the observed electric field orientation varies somewhat over the length of the NTFs, though in all regions, the distribution is very well ordered. It is difficult to interpret these results, however, since no Faraday correction has been applied. For other GC NTFs, where Faraday corrections have been made, intrinsic magnetic field orientations are observed to be primarily parallel to the extent of the NTFs \citep{Lang1999a, Lang1999b,YWP1997,Gray1995}, consistent with the idea that the NTFs are tracing a well-ordered magnetic field in the plane of the sky and presumably at the GC. 

As described in Section \ref{sec:mag}, Figure \ref{fig:vec_over_x} shows the Faraday-corrected vectors that represent the intrinsic magnetic field of the region observed in the Arc NTFs. The intrinsic magnetic field of Figure \ref{fig:vec_over_x} reveals regions where the magnetic field is parallel to the extent of the Arc NTFs. However, there are other regions where the magnetic field is rotated from that parallel orientation to a well-ordered orientation that is primarily perpendicular to the Arc NTFs. These regions of well-ordered (but not parallel to the NTFs) instrinic magnetic field orientation are puzzling. It is possible that a foreground magnetic field, unrelated to that traced by the Arc NTFs, is superimposed in our field of view. It could be this foreground field is responsible for the regions of rotated magnetic field orientation. A possible clue could come from the locations of the elongated polarized structures discussed in Section \ref{sec:pol_elong} relative to the different magnetic field orientations. The elongated polarized features appear to be at points where the magnetic field alternates from a parallel to a rotated orientation (Section \ref{sec:mag}). These polarized features could represent components of a foreground structure that possesses an ordered magnetic field that is rotated with respect to the Arc NTFs, thus producing the regions of differently oriented magnetic field we observe. 

Alternatively, the unusual magnetic field pattern could be a result of an intrinsic property of the Arc NTFs. \citet{YM1987} find evidence of the Arc NTFs twisting about each other in their observations and postulate that this twisting could generate a rotated magnetic field orientation. In our total intensity images of the Arc NTFs (see Figure \ref{fig:tot_I_4}), we do not see any evidence of twisting in the NTFs meaning it is unlikely to be the Arc NTFs that are producing the rotated magnetic field regions.

\subsection{Structures Local to the Radio Arc} \label{sec:helix}

\begin{figure*}[tp]
    \gridline{\fig{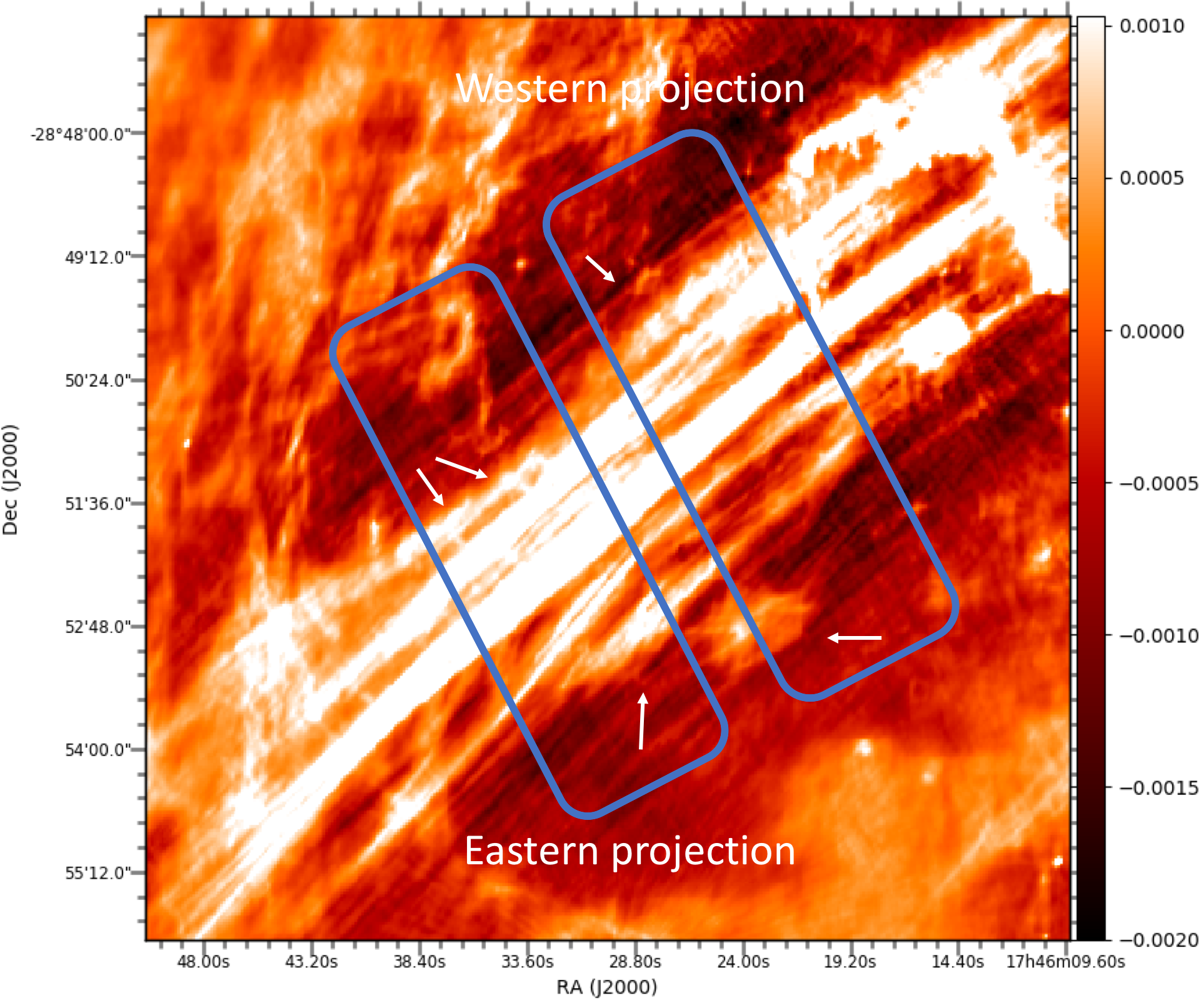}{1.0\textwidth}{}}
    \caption{5 GHz total intensity distribution with estimated projections for the helical segments overlayed as blue outlines. White arrows mark total intensity features discussed in the text. Labels for the wedges follow the naming convention employed in the text. The color bar is in units of Jy beam$\rm^{-1}$.}
    \label{fig:cross_1}
\end{figure*}

\begin{figure*}
    \gridline{\fig{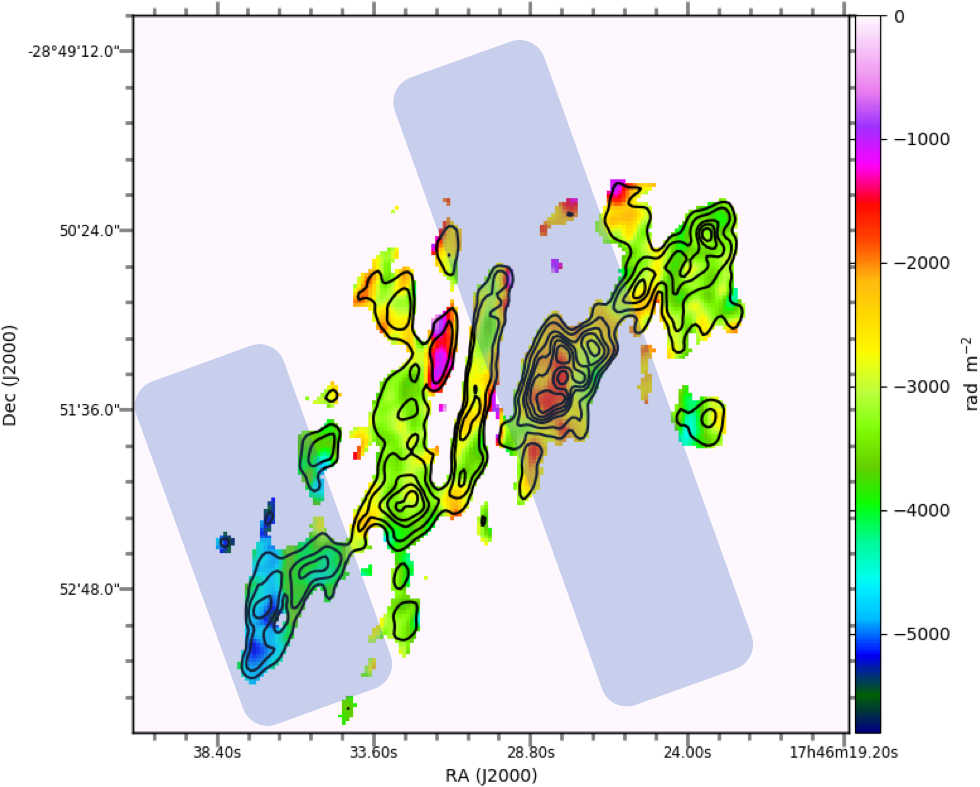}{1.0\textwidth}{}}
    \caption{10 GHz RM distribution in color scale with polarized intensity contours shown in black. Contour levels are at 0.2, 0.4, 0.6, 0.7, 0.8, and 0.9 times the maximum polarized intensity value of 23.2 mJy beam$\rm^{-1}$. The blue wedges represent regions of the helical segments shown in Figure \ref{fig:cross_1} that overlap the polarized intensity emission.}
    \label{fig:cross_2}
\end{figure*}

\begin{figure*}[tp]
    \gridline{\fig{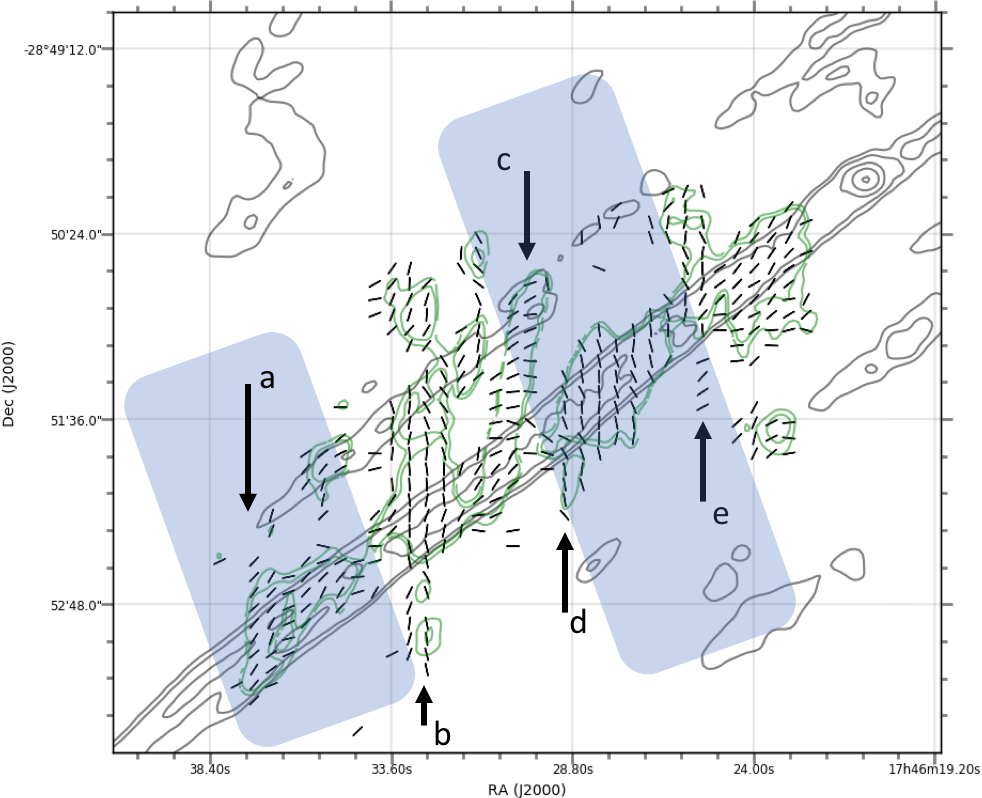}{1.0\textwidth}{}}
    \caption{10 GHz total intensity contours in black, 10 GHz polarized intensity contours in green, and 10 GHz intrinsic magnetic field orientations. The total intensity contours are at 1, 2, 4, 8, 16, and 32 times the rms noise level of 2.43 mJy beam$\rm^{-1}$ and the polarized intensity contours are at 0.2 and 0.3 times the maximum polarized intensity value of 23.2 mJy beam$\rm^{-1}$. The elongated structures previously discussed in Sections \ref{sec:pol_elong} and \ref{sec:thorn_disc} are labeled in the same manner employed in the upper panel of Figure \ref{fig:ps}. The blue wedges represent regions of the helical segments shown in Figure \ref{fig:cross_1} that overlap the polarized intensity emission.}
    \label{fig:cross_4}
\end{figure*}

\begin{figure*}
    \gridline{\fig{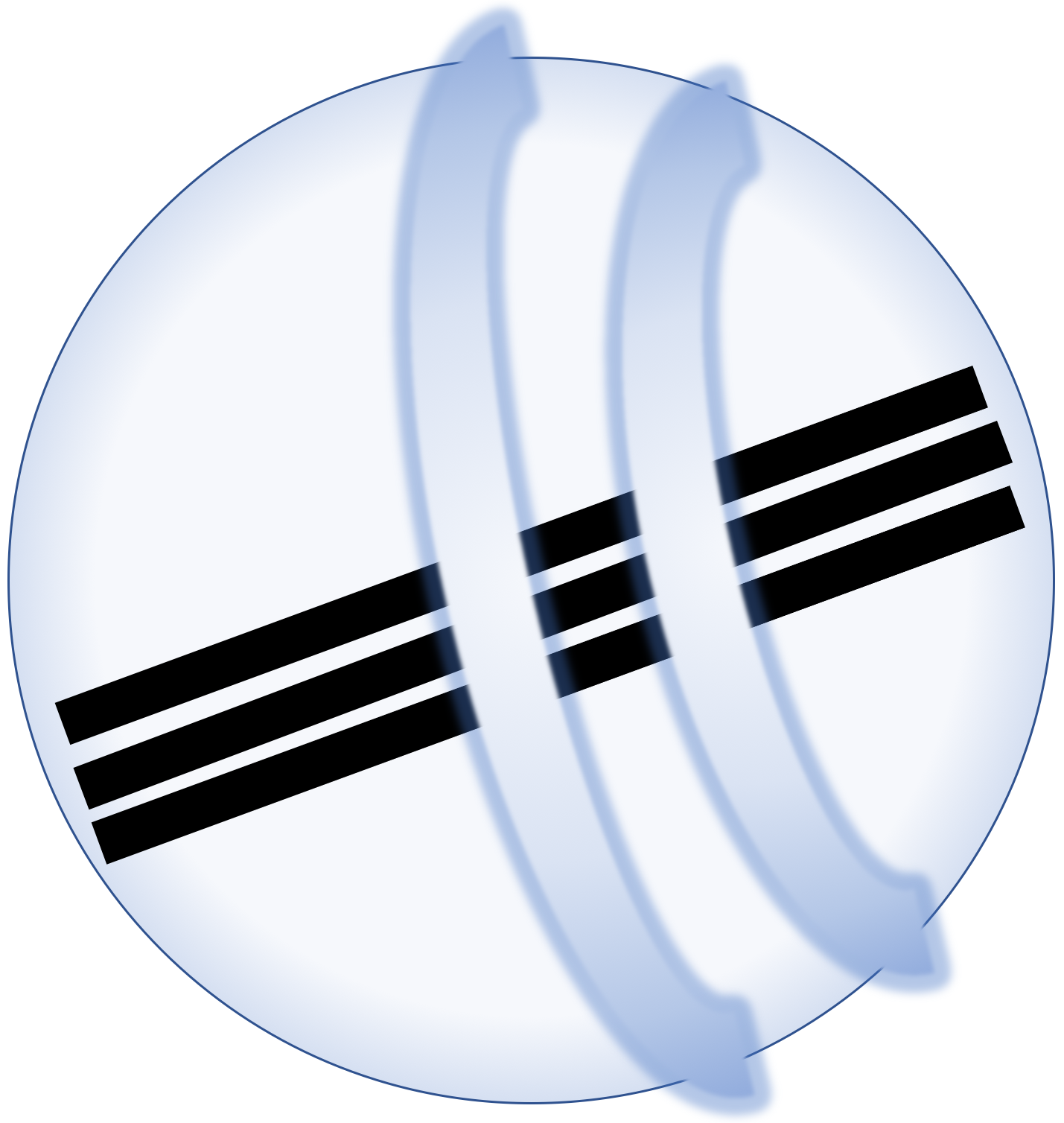}{0.3\textwidth}{a)}
              \fig{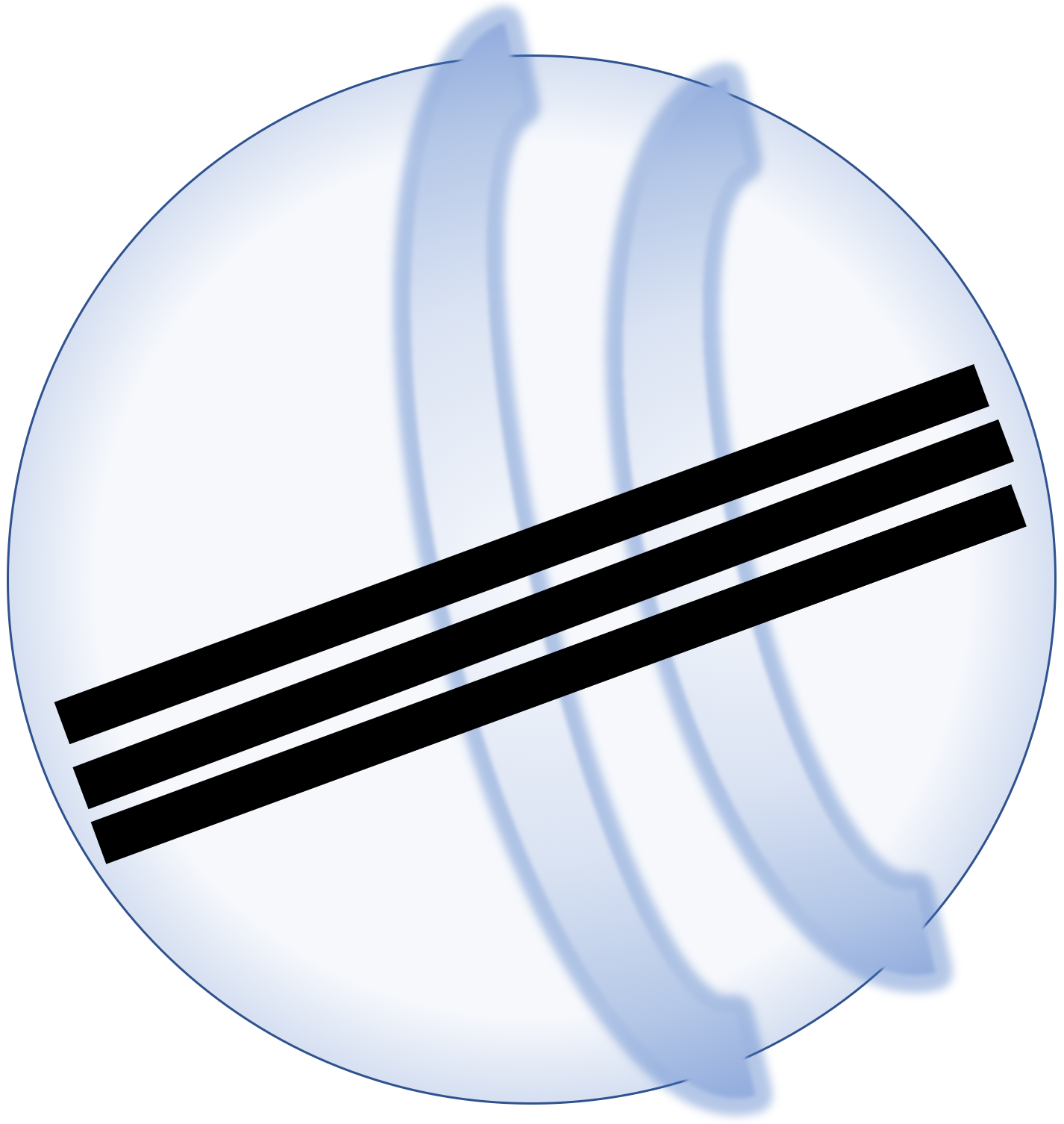}{0.3\textwidth}{b)}
              \fig{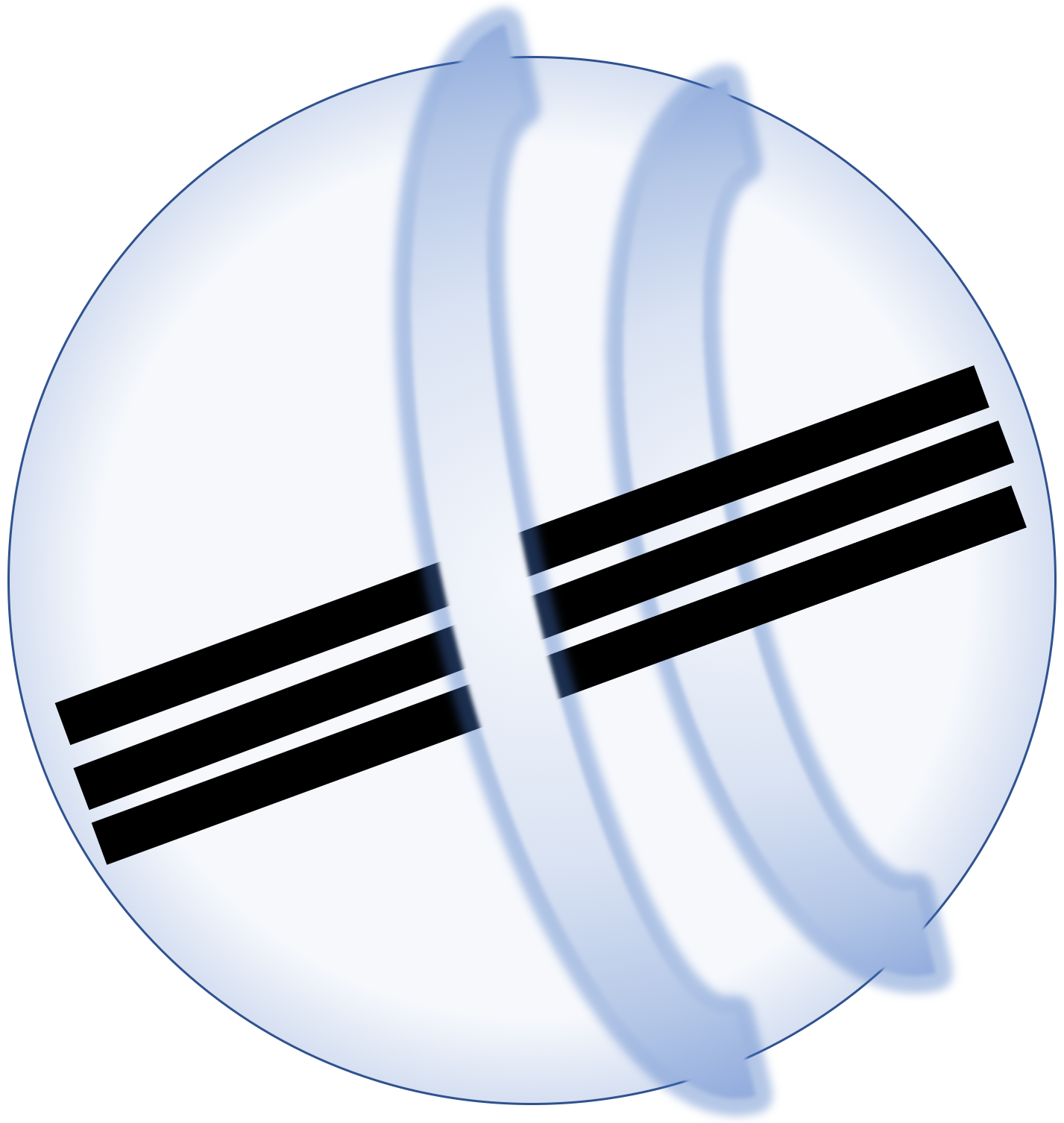}{0.3\textwidth}{c)}}
    \caption{Sketches detailing the physical relationship between the Arc NTFs, the radio bubble, and the helical segments. Panels a-c show situations where the helical segments are foreground, background, or alternating in relative line-of-sight orientation with respect to the Arc NTFs (black lines).}
    \label{fig:toy}
\end{figure*}

As described above, the polarization and magnetic field orientation in the Arc NTFs appear to be more complex than in the other GC NTFs as shown in Figures \ref{fig:ps} and \ref{fig:vec_over_x}. One suggestion to explore is that the unique environment in which the Arc NTFs are located may contribute to these more complicated observed properties. In particular, several large-scale and diffuse helical segments are observed to approach the Arc NTFs in Figure \ref{fig:tot_I_1}. The presence and possible superposition of these helical segments with the Arc NTFs could affect the appearance of the received emission from the Arc NTFs. As observed in Figures \ref{fig:tot_I_1}, \ref{fig:tot_I_2} and \ref{fig:tot_I_3}, the helical segments approach the Northern edge of the NTFs and are extremely diffuse in nature. As they intersect the Arc NTFs they appear to fade to the noise levels of the total intensity distributions. It is possible to make an estimate of where these helical segments may cross the Arc NTFs. Our estimates are shown in Figure \ref{fig:cross_1} superimposed on the 5 GHz total intensity image (Figure \ref{fig:tot_I_2}). The locations chosen for these estimates were based on the locations of the helical segments and the locations of faint total intensity features which have been marked with white arrows in Figure \ref{fig:cross_1}, which are possible extensions of the helical segments. In the following sections, we assess whether the possible presence of helical segments along the line of sight toward regions of the Arc NTFs can explain some of the puzzling features we observe in our data of the Arc NTFs.

\subsubsection{Helical Segments and Polarized Intensity}
Figure \ref{fig:cross_2} shows the 10 GHz RM distribution with the polarized intensity contours in black. While the eastern helical segment approaches the Arc NTFs at a region of low polarized intensity of 5.3 mJy beam$\rm^{-1}$ (vs. a maximum polarized intensity value of 23.2 mJy beam$\rm^{-1}$), the western helical segment approaches the point with the peak polarized intensity value. Furthermore, while the distribution of polarized intensity is sparse for the eastern segment, the western segment covers a far more extensive distribution of polarized intensity. 

\subsubsection{Helical Segments and RM Distribution}
Turning to the RM distribution shown in Figure \ref{fig:cross_2}, the eastern segment coincides with RM magnitudes generally in the range of 3500 - 4500 $\rm rad\,\, m^{-2}$ whereas the western segment passes RM magnitudes generally within the range of 1500 - 2500 $\rm rad\,\, m^{-2}$. The RMs in the eastern segment are generally larger than those seen for the western segment, and furthermore the eastern segment coincides with locations of maximum RM magnitude.

\subsubsection{Helical Segments and Magnetic Field Orientation}
Figure \ref{fig:cross_4} elucidates the magnetic field behavior at the regions where the segments approach the Arc NTFs. The eastern segment passes over a magnetic field that traces the extent of the Arc NTFs (coinciding with the left-most rectangular region marked in Figure \ref{fig:vec_over_x}). However, the western segment coincides with a magnetic field predominantly rotated with respect to the Arc NTFs even in locations of high total intensity (see Figure \ref{fig:cross_4}, coinciding with the right elliptical region of Figure \ref{fig:vec_over_x}).

We have seen in Sections \ref{sec:mag} and \ref{sec:mag_disc} how the elongated structures generally coincide with transition regions between parallel and rotated magnetic field regions. Figure \ref{fig:cross_4} marks the locations of the polarized elongated structures  and all of them are at least partially coincident with the helical segments. The spatial coincidence of the elongated polarized features and the helical segments could mean they are all components of the same large-scale structure. This structure is seemingly external to the Arc NTFs and, if magnetized, could be responsible for the regions of rotated magnetic field we observe.

As seen in Figures \ref{fig:cross_2} and \ref{fig:cross_4}, the eastern and western helical segments are coincident with locations of the Arc NTFs that display different polarization, RM, and magnetic field characteristics. We would now like to determine a possible geometry of the helical segments based on their effects on the distributions of the Arc NTFs.

\subsubsection{Sketch of Helical Segment Arrangement}
We develop a set of sketches, shown in Figure \ref{fig:toy}, to motivate our analysis of the physical relationship between the helical segments and the Arc NTFs. In these sketches, we assume that the helical segments are diffuse and low-intensity parts of a larger structure, such as the radio bubble identified in Section \ref{sec:tot_res}, that surrounds the Arc NTFs. The helical segments could be components of the radio bubble since the total intensity images reveal that the helical segments seem to connect to this enveloping structure (Figure \ref{fig:tot_I_1}). These helical segments could be density fluctuations in the radio bubble, representing regions of elevated electron density. They are also only faintly illuminated, like the radio bubble in general, and so they fade from detection in Figures \ref{fig:tot_I_1} - \ref{fig:tot_I_3} when viewed directly face-on where they would cross the Arc NTFs. The three sample situations shown in Figure \ref{fig:toy} depict a region where (a) the helical segments cross foreground to the Arc NTFs, (b) the helical segments cross background to the Arc NTFs, and (c) the helical segments alternate being foreground and background to the Arc NTFs. We can assess which of these possible situations is most reflective of reality by comparing them to the effects the helical segments have on the Arc NTF data sets.

If all the helical segments were foreground, as shown in panel a, then we would expect the two helical segments we analyze to have similar depolarizing effects on the polarized intensity distribution of the Arc NTFs. However, only the eastern segment corresponds with a decrease in polarized intensity, and so we consider it unlikely that both segments are foreground. This assessment rules out the geometry shown in panel a. Similar arguments can be used to rule out the geometry shown in panel b. The geometry in panel c, which shows the helical segments alternating between foreground and background is more akin to the behavior we observe for the eastern and western segment, and so we consider this situation to be an accurate representation of reality.

In the sketch of panel c we depict the eastern helical segment as a foreground structure and the western segment as being background. However, since we do not know the sign of the line of sight magnetic field in these helical segments, we are unable to determine whether a foreground segment would increase or decrease the RM obtained for lines of sight passing through it. We may be able to determine in more detail which of these segments is foreground in a subsequent paper utilizing more sophisticated RM analysis.
	
\section{CONCLUSIONS} \label{sec:conc}

In this paper we have conducted a comprehensive study of the Arc NTFs utilizing the upgraded polarimetric capabilities of the VLA to obtain sensitive total and polarized intensity distributions as well as RM and intrinsic magnetic field distributions. We summarize the key findings of this work here:

\begin{enumerate}
    \item Our total intensity images reveal a complex region surrounding the Arc NTFs consisting of point and compact sources, a radio lobe surrounding the Arc NTFs, multiple HII regions (notably the Sickle and Pistol), helical segments, and the Arc NTFs themselves. We find evidence for additional subfilamentation within the Arc NTFs beyond what has previously been observed, with individual NTFs having narrowest widths of about 0.2 pc.
    \item Our polarized intensity results reveal a patchy distribution, which agrees with previous observations of the Arc NTFs \citep{YM1987}. The distribution is confined within a 4\arcmin~region in the center of the NTF cluster. We confirm the existence of the `thorn' features originally seen in \citet{Inoue1989}. We identify additional elongated structures and determine that these structures are likely local to the Arc NTFs or slightly foreground to them. We observe significant depolarization between the 10 GHz and 5 GHz data sets, the source of which is unclear.
    \item For the first time we reveal the RM distribution towards the Arc NTFs. The RM values are negative towards the Arc NTFs with values ranging from -500 to -5500 rad m$\rm^{-2}$. These RM measurements were produced using hundreds of frequency channels, and they agree with previous RM measurements of the GC NTF population.
    \item We find that the rotation we observe is likely due to external Faraday effects, although we are unable to determine the detailed Faraday screen arrangement with this analysis. We make the initial prediction that rotation occurs in the GC region, characterized by the scattering medium described in \citet{Lazio1998b}. 
    \item Our constrained RM distribution allows us to produce the intrinsic magnetic field distribution for the Arc NTFs. The intrinsic magnetic field alternates between being aligned along the extent of the Arc NTFs and rotated with respect to the extended axis of the Arc NTFs. This makes quite a contrast to the magnetic field distributions obtained for other NTFs which show regular magnetic fields that closely trace the extent of the NTFs. The regions of rotated field are bordered by the polarized elongated structures, so we postulate that the regions of rotated field could be an external magnetic field system traced by these elongated features.
    \item We confirm the presence of total intensity helical segments which cross the Arc NTFs, as originally observed in \citet{YM1987}. We find an alternating pattern of foreground to background with respect to the Arc NTFs for these helical segments.
\end{enumerate}

\acknowledgements This material is based upon work supported by the National Science Foundation under Grant No. AST-1615375. The National Radio Astronomy Observatory is a facility of the National Science Foundation operated under cooperative agreement by Associated Universities, Inc. D.P. would like to thank Arran Gross and Josh Steffen for helpful comments on the analysis of the data. D.P. and C.C.L. would like to thank Genna Crom for the helpful assistance she provided for the polarized intensity sections of the paper and James Toomey for the calibration of the data.

\software{APLpy \citep{Robitaille2012},
    Astropy \citep{Greenfield2014},
    CASA \citep{McMullin2007},
    LMFIT \citep{Newville2016},
    Matplotlib \citep{Hunter2007}
    }

\bibliographystyle{aasjournal}
\bibliography{astronomy}
	
\end{document}